\documentstyle[12pt,fleqn,twoside] {report}
\topmargin
\headheight
  \title{{\sc Cosmic ray acceleration at relativistic and ultrarelativistic
shock waves}}
\author{
{\Large Janusz Bednarz} \\~\\
{\it Jagiellonian University Astronomical Observatory}\\
~\\~\\~\\~\\~
  Ph.D. thesis\\
  written under the supervision of\\
  Dr. hab. Micha{\l} Ostrowski \\~\\~\\~\\~\\~\\~\\~\\~\\~\\~\\
  Cracow \\ November 1999} 
\date{~}
 
\begin{document}
 \maketitle
 \begin{quote}
~\\~\\~\\~\\~\\~\\~\\
ACKNOWLEDGMENTS\\
\vspace*{-0.4truecm}

{
\hoffset -0.93cm
\it I am deeply grateful to the supervisor
of my thesis Dr. hab. Micha{\l} Ostrowski for his guidance and
support and, the advice of a number of colleagues and collaborators
including Krzysztof Chy\.{z}y, W{\l}odzimierz God{\l}owski,
Zdzis{\l}aw Golda, Jacek Guzik, Marek Jamrozy, Jacek Knapik,
Tomasz Kobak, Tomasz Kundera, Krzysztof Ma\'slanka, Adam Michalec, Grzegorz
Micha{\l}ek, Katarzyna Otmianowska-Mazur, Stanis{\l}aw Ry\'s,
Gra\.{z}yna Siemieniec-Ozi\c{e}b{\l}o, Marian Soida, Marek
Szyd{\l}owski, Bogdan Wszo{\l}ek and Stanis{\l}aw Zo{\l}a from the
Cracow Astronomical Observatory were helpful during my work.
The presented computations were partly done on a CONVEX Exemplar
SPP1000/XA-16 and HP Exemplar S2000 at ACK `CYFRONET' in Krak\'ow.
Several parts of the work were supported by} \rm Komitet Bada\'{n}
Naukowych {\it through  grants PB~1117/P03/94/06, PB~179/P03/96/11
and PB~258/P03/99/17.}~~
\end{quote}

\hoffset -0.93cm

\newpage
\tableofcontents
\newpage
\chapter{Introduction}

Ever since the discovery of radiation which comes from cosmos by Hess in 1912
and christened by Millikan in 1925 as `cosmic rays', physicists and astronomers
have speculated upon their origin. Fermi (1949) made the first serious attempt
at explaining the power law nature of the cosmic ray spectrum. He noted that
a particle could increase its energy at collisions against magnetic field
irregularities. In his model cosmic rays interact with galactic molecular clouds
that move randomly. Particles increase their energy in head-on collisions which
are more frequent than overtaking collisions when they loose energy. The process
is known as the second-order Fermi acceleration because the mean particle
momentum gain $\Delta p/p$ in one interaction is proportional to $(V/v)^{2}$,
$V$ is the root-mean-square velocity of a cloud and $v$ is the particle velocity,
considered below to be comparable to the speed of light -- c. Presently the
second-order Fermi acceleration is considered in plasma where the magnetic field
fluctuations play a role of the Fermi `clouds'.

\noindent
{\bf Nonrelativistic shocks.} The concept that shock waves accelerate particles
in a mechanism similar to the one described by Fermi (1949) appeared in four
seminal papers: Krymski (1977), Axford et al. (1977), Bell (1978a,b) and
Blandford \& Ostriker (1978). The idea was foreshadowed by Hoyle (1960) who
postulated that shocks could efficiently accelerate particles but without
specifying a mechanism. Parker (1958) and Hudson (1965, 1967) attempted to
obtain such mechanism based on pairs of converging shocks and, most notably,
Schatzman (1963) constructed a theory based on perpendicular shocks. Contrary
to the original mechanism in the convergent shock flow pattern particles interact
with the flowing plasma only like in head-on collisions. Mean momentum gain
in such interaction is proportional to $U_{1}/c$ ($U_{1}$ is the shock velocity)
and hence the process is known as the first-order Fermi acceleration. Efficiency
of the first-order relative to the second-order Fermi acceleration equals
roughly $U_{1}/V_{A}$, where $V_{A}$ is the Alfv\'en speed in plasma
(cf. Ostrowski \& Schlickeiser 1996).

A shock wave, or briefly a shock, can be described as a sharp transition layer
which propagates through plasma with a velocity exceeding the speed of sound and
changes its state through the compression. The thickness of the layer is
determined by the physical process responsible for thermodynamic parameters
transfer from incoming plasma, upstream of the shock, to flowing away plasma,
downstream of the shock. In tenuous plasma the transfer proceeds through
collective electromagnetic effects and the shock width is of the order of the
gyroradius of a thermal ion. In the acceleration process we will consider
relativistic particles which move with speeds close to the speed of light and
have a gyroradii much larger than thermal ions and consequently they see the
shock as a discontinuity.

The acceleration processes in nonrelativistic -- $U_{1}\ll c$ -- shocks
yield power-law particles momentum spectra, $f(p)\propto p^{-\alpha}$, with a
very simple formula for the spectral index of accelerated particles
$$
\alpha = \frac{3r}{r-1} \qquad ,\eqno(1.1)
$$
where
$$
r = \frac{\gamma_{a}+1}{\gamma_{a}-1+2 M^{-2}} \eqno(1.2)
$$
is a shock compression ratio, $M$ is the shock Mach number and $\gamma_{a}$ is
the plasma adiabatic index. For a strong shock, $M\rightarrow\infty$, propagating
in a nonrelativistic plasma with $\gamma_{a} = 5/3$ we have $r\rightarrow 4-$
and $\alpha\rightarrow 4+$. This is encouragingly close to the index of 4.3
inferred for the source of the galactic cosmic rays. Similarly, the acceleration
time expressed by a simple diffusive formula is discussed in Section 3.

\noindent
{\bf Relativistic shocks.} A consistent method to tackle the problem of
first-order Fermi acceleration in relativistic shock waves was conceived by
Kirk \& Schneider (1987a; see also Kirk 1988). They assumed a parallel shock
geometry and that particles are subject to pitch-angle scattering on each
side of the shock. By extending the diffusion approximation to higher order
terms in the anisotropy of the particle distribution, they obtained solutions
to a kinetic  equation of the Fokker--Planck type with the isotropic form of
pitch  angle diffusion coefficient. Since pitch-angle scattering conserves the
particle momentum in the fluid frame, the energy spectrum is obtained by matching
the solutions at the shock. Their $Q_{L}$ method yielded a particle
energy spectral index for strong nonrelativistic shocks as $\sigma\simeq 2.0$
-- where $\sigma \equiv \alpha - 2$ -- in agreement with previous results.
For relativistic shocks with realistic compression of Heavens \& Drury (1988),
the method produced particle spectra with $\sigma$ slightly smaller than 2
provided the Lorentz factor of the shock $\gamma\leq 5$, and slightly larger
at higher $\gamma$. The authors derived also an angular distribution function
at the shock as measured in the upstream and the downstream fluid frame.
In the upstream fluid frame the distribution is strongly peaked even for
a mildly relativistic case of $U_{1}=0.3c$.

Next, Kirk \& Schneider (1988) extended the analysis by
involving  both diffusion and large-angle scattering in particle  pitch
angle. They discovered that -- in relativistic shock waves -- the
presence of large angle scattering can substantially modify the spectrum of
accelerated particles. An extension of the Kirk \& Schneider's (1987a)
approach to more general conditions in the shock was given by Heavens \&
Drury (1988) who took into consideration the fluid dynamics of
relativistic shock waves. They also noted that the resulting particle
spectral indices depend on the perturbations spectrum near the shock in
contrast to the nonrelativistic case.

Kirk \& Heavens (1989) considered the acceleration process in shocks with
magnetic fields oblique to the shock normal (see also Ballard \& Heavens
1991) by extending the method of Kirk \& Schneider (1987a). Oblique shock
fronts may be conveniently classified into two categories: subluminal and
superluminal. In the former ones it is possible to find a Lorentz
transformation to a frame of reference in which the electric field is zero
in both the upstream and the downstream regions, and the shock front is
stationary. In this frame, called the de Hoffman-Teller frame (de Hoffman
\& Teller 1950), the energy of a particle remains constant provided it does
not suffer scattering. Superluminal shocks, however, do not admit a
transformation to such a frame of reference. They correspond to shock
fronts in which the point of intersection of the front with a magnetic
field line moves at a speed greater than c. Kirk \& Heavens used the de
Hoffman-Teller frame to consider the subluminal shocks. They showed,
contrary to nonrelativistic results again, that such shocks led to flatter
spectra than parallel ones approaching the value $\sigma\simeq 1.0$ when
the shock velocity along the magnetic field $U_{B}\simeq c$. Their work
relied on the assumption of adiabatic invariant $p_\perp^2 /B$ conservation
for particles interacting with the shock, which restricted considerations
to the case of a nearly uniform magnetic field upstream and downstream of
the shock.

A different approach to particle acceleration was presented by
Begelman \& Kirk (1990) who noted that in relativistic shocks most
field configurations lead to super-luminal conditions for the
acceleration process. In such conditions, particles are accelerated in a
single shock transmission by drifting parallel to the electric field
present in the shock. Begelman \& Kirk showed that there is more
efficient acceleration in relativistic conditions than that predicted by
a simple adiabatic theory.

The acceleration process in the presence of finite amplitude magnetic
field perturbations was considered by Ostrowski (1991; 1993) and Ballard
\& Heavens (1992). Ostrowski considered a particle acceleration process
in the relativistic shocks with oblique magnetic fields in the presence
of field perturbations, where the assumption $p_\perp^2 /B  = {\rm const}$
was no longer valid. To derive particle spectral indices he used a method
of particle Monte Carlo simulations and noted that the spectral index was
not a monotonic function of the perturbation amplitude enabling the
occurrence of steeper spectra than those for the limits of small and large
perturbations. It was also revealed that conditions leading to very flat
spectra involve an energetic particle density jump at the shock. The
acceleration process in the case of a perpendicular shock shows a
transition between the compressive acceleration described by Begelman \&
Kirk (1990) and, for larger perturbations, the regime allowing for
formation of the wide range power-law spectrum. The Ostrowski (1991)
method was based on the `mean field + perturbation' decomposition of
magnetic field, i.e. a particle is considered to propagate in the mean
field along its undisturbed `adiabatic' trajectory, while the magnetic
field inhomogeneities are allowed for by perturbing the trajectory
parameters in finite time-steps. The simulations of Ostrowski (1993) were
based on the numerical integration of the particle equations of motion in
a perturbed magnetic field. Finite-amplitude field perturbations were
described with analytic formulae as a superposition of static sinusoidal
waves.

The analogous simulations by Ballard \& Heavens (1992) for highly
disordered background magnetic fields show systematically steeper
spectra in comparison to the above results, as discussed by Ostrowski
(1993). In terms of the Lorentz factor of the shock Ballard \& Heavens
found a rough relation $\alpha \simeq (3\gamma +1)/8$ that is valid
up to $\gamma \simeq 5$. They checked their results considering
different power-law fluctuations spectra for the magnetic field 
and stated that differences between the resulting particle spectra were
quite small.

The particle spectrum formation in the presence of non-linear coupling
of accelerated particles to the plasma flow has been commented by
Ostrowski (1994).

The shock waves propagating with relativistic velocities rise also
interesting questions concerning the cosmic ray acceleration time
scale, $T_{acc}$. Until our results published in 1996 (Bednarz \&
Ostrowski 1996 - see chapter 3) there was only somewhat superficial
information available about that problem. A simple comparison to the
nonrelativistic formula based on numerical simulations shows that
$T_{acc}$ relatively decreases with increasing shock velocity for
parallel (Quenby \& Lieu 1989; Ellison et al. 1990) and oblique
(Takahara \& Terasawa 1990; Newman et al. 1992; Lieu et al. 1994;
Quenby \& Drolias 1995; Naito \& Takahara 1995) shocks. However, the
numerical approaches used there, based on assuming the particle
isotropization at each scattering, neglect or underestimate a
significant factor controlling the acceleration process -- the
particle anisotropy. Ellison et al. (1990) and Naito \& Takahara
(1995) included also derivations applying the pitch angle diffusion
approach. The calculations of Ellison et al. for parallel shocks show
similar results to the ones they obtained with large amplitude
scattering. In their computations for the shock with velocity $0.98\,c$
the acceleration time scale is reduced on a factor $\sim 3$ with
respect to the nonrelativistic formula. Naito \& Takahara considered
shocks with oblique magnetic fields. They confirmed the reduction of
the acceleration time scale with increasing inclination of the magnetic
field derived earlier for nonrelativistic shocks (Ostrowski 1988).
However, their approach neglected the effects of particle cross field
diffusion and assumed the adiabatic invariant conservation at particle
interactions with the shock. These two simplifications limit their
results to the cases with small amplitude turbulence near the
shock\footnote{One should note that the spatial distributions near the
shock derived by these authors (their figures 1 and 2) do not show a
particle density jump proved to exist in oblique relativistic shocks by
Ostrowski (1991). It is also implicitly present in analytic derivations
of Kirk \& Heavens (1989).}. One should also note that comparing some
of the mentioned papers the derived time scales to the nonrelativistic
expression does not have any clear physical meaning when dealing with
relativistic shocks.

In the present paper we use pitch angle diffusion approximation for
particle transport in the acceleration process. Let us note that some
earlier derivations of the acceleration time scale were based on the
numerical simulations involving particle scattering at point like
scattering centers isotropizing the particle momentum at each
scattering, the so called large angle scattering model. This approach
does not provide a proper description for the acceleration processes
in shock waves moving with velocities comparable to the particle
velocity because it removes particle anisotropy and changes the factors
related to it. Moreover, against arguments presented in some papers
such scattering patterns can not be realized in turbulent magnetic
fields near relativistic shocks, where most particles active in the
acceleration process are able to diffuse only a short distance below
a few particle gyroradii off the shock\footnote{ However, for the
nonrelativistic shock velocity and particles much above the injection
energy such approximations can be safely used (cf. Jones \& Ellison
1991).}. Such distances are most often insufficient to allow for
big particle pitch-angle changes. In shocks with oblique magnetic fields
such large angle scattering patterns can substantially change the shape
of the accelerated particle spectrum with respect to the pitch angle
diffusion model. Additionally, as an individual particle interaction
with the shock can involve a few revolutions along the magnetic field,
the usually assumed adiabatic invariant conservation, $p_\perp^2/B =
{\rm const}$, cannot be valid for short inter-scattering intervals.

\noindent
{\bf Ultrarelativistic shocks.}
The acceleration mechanism described in section 4.1 is quite different
from that in the nonrelativistic and mildly relativistic regime so that
we distinguish a class of ultrarelativistic shocks if their Lorentz
factors $\gamma\gg1$. The condition $\gamma\gg1$ implies also some
simplifications that allow to consider ultrarelativistic shocks as a
separate class.
First, the magnetic field inclination downstream of the shock is, in
practice, always perpendicular to the shock normal as one can derive
from Eq. 2.14. Similarly, we can approximate in Eq. 2.13 the ratio of
the value the magnetic field downstream of the shock to upstream as
$B_{2}/B_{1}\simeq \sqrt{8} \gamma\sin\psi_{1}$. A turbulence
downstream of the shock could amplify this value and for example
assuming equipartition with the thermal pressure downstream, one
obtains $B_{2}/B_{1}\sim (c/V_{A})\gamma$. Moreover, independently
of the plasma composition (proton-electron or electron-positron) the
shock velocity relative to the downstream medium is $U_{2}=c/3$ in
the limit of large $\gamma$.

The ultrarelativistic shocks are characterized by large
anisotropy of particle momentum distribution near the shock that
was presented in Bednarz \& Ostrowski (1998, see Figs.~4.4~-~4.7
below). The values of two main parameters describing the acceleration
process, namely the energy spectral index and the acceleration time,
are independent of shock conditions if fluctuations upstream of the
shock ensure the acceleration process to be effective. They tend to
2.2 (spectral index, Bednarz \& Ostrowski 1998; also Bednarz \&
Ostrowski 1997a,b) and 1.0 $r_{g}/c$ (acceleration time, Bednarz 1998,
1999). The rough analytical calculations
of Gallant \& Achterberg (1999) are consistent with the Bednarz \&
Ostrowski (1998) paper and Gallant et al. (1998) confirm the value
of spectral index for the specific condition of the extremely
disordered magnetic field downstream of the shock.

Ultrarelativistic shocks are considered as sources of cosmic rays
with energies exceeding $10^{20}$ eV and several papers suggested
that gamma ray bursts (GRBs) could be sources of these particles
(cf. Waxman 1995, Vietri 1995). Vietri (1995) argued that in the
Fermi-type acceleration at an ultrarelativistic shock, a particle
could have an relative energy gain $\sim\gamma^{2}$ per shock crossing
cycle. Gallant \& Achterberg (1999) showed that particles with initial
momenta isotropically distributed upstream of the shock gain
$\sim\gamma^{2}$ energy, but only at the first interaction of the shock.
They also showed that for parameters typical of the millisecond pulsars in
the neutron star binaries observed in our Galaxy, the gamma ray burst
blast wave would decelerate within the pulsar wind bubble, yielding an
energy spectrum $\sigma\simeq 2$ for the boosted particles. Moreover,
this spectrum would typically extend over the energy region
$10^{18.5}-10^{20}$ eV, i.e. precisely where the ultra-high-energy
cosmic rays (UHECR) component is observed. Bednarz (1999) suggested
that such extremely energetic particles could be produced by
reflections of the shock directly in GRBs.

\noindent
{\bf Relativistic shocks in astrophysical objects.}
Results presented further in the theses could be applied in models
of some galactic and extragalactic objects. One of those are active
galactic nuclei where a central black hole ejects plasma in the form
of relativistic jets. A few tens of blazars has been detected in GeV
$\gamma$-rays by the EGRET detector (von Montigny et. al. 1995). It
is widely believed that the $\gamma$-ray production in blazars is
strictly related to the existence of relativistic jets because many
of them show superluminal motions (Vermeulen \& Cohen 1994). 
Jiang et al. (1998) applied the K\"onigl inhomogeneous jet model
(Blandford \& K\"onigl 1979; K\"onigl 1981) to a sample of quasars
and BL Lacs objects and found the Lorentz factors of jets to be
a significant part the ultrarelativistic ones. In unified schemes
for active galactic nuclei the Fanaroff-Riley type II (FR II) radio
sources are formed by AGNs, similarly to blazars, but jets are ejected
at higher angles to the line of sight. Evidence that they are
relativistic even on tens or hundreds kiloparsec scales suggest that
the hotspots in these sources are the downstream regions of
relativistic shocks.

The recent finding of microquasars in our Galaxy, a class of objects
that mimic -- on scales million of times smaller -- the properties
of quasars opened new possibilities to study physical processes
in accretion disks of black holes. The observations of Mirabell
\& Rodriguez (1994), Tingay et al. (1995), and Hjellming \& Rupen (1995)
confirm the existence of relativistic flows related to these objects,
and it is expected that they form relativistic shocks in the
interstellar medium.

A relativistic wind of magnetized electron-positron plasma blowing
from a pulsar with the flow Lorentz factor of $\sim 10^{6}$ is
expected to form a termination shock (e.g. Kennel \& Coroniti 1984;
Gallant \& Arons 1994 and Chiueh et al. 1998). Non-thermal radiation
apparently seen in the class of such objects -- plerions -- suggests
the existence of acceleration processes inside the nebula. The Crab
Nebula as the young and energetic source is the best plerion to study
it. Recent optical observations of Crab Nebula using the Hubble Space
Telescope and also the X-ray observations of ROSAT (cf. Hester et al.
1995) show a fascinating structure of jets, a torus of X-ray emission
and complexes of sharp wisps. $\gamma$ ray observations of the Crab
Nebula exhibit the existence of extremely energetic electrons near
the pulsar (cf. de Jager et al. 1996). The electron energy is a few
magnitudes larger than that in the blowing wind so an acceleration
mechanism has to take place near the pulsar. Gallant \& Arons (1994)
proposed a mechanism where electrons gain their energy from
electromagnetic waves generated by gyrating ions. The mechanism tries
to explain wisps at the distance of 10" from the pulsar but a knot
found at 0.7" (cf. Hester et al. 1995) is not explained in the model.
We expect that acceleration mechanism presented by Bednarz \& Ostrowski
(1998) and Bednarz (1999) is able to account for the generation of such
energetic electrons at if the ultrarelativistic shock formed near the
pulsar.

Observations carried out by the Burst and Transient Source Experiment
show that GRBs originate from cosmological sources (Meegan et al. 1992
and Dermer 1992). Identification of the host galaxy for the GRB 971214
(Kulkarni et al. 1998) and several other bursts causes there is little
doubt now that some, and most likely all GRBs are cosmological. These
phenomena are surely related to ultrarelativistic shocks with
$\gamma>10^{2}$ (cf. Woods \& Loeb 1995). The power-law form of the
spectrum often observed at high photon energies suggests the existence
of nonthermal populations of energetic particles. Bednarz \& Ostrowski
(1998, see chapter 4 below) showed that such shocks are able to
accelerate charged particles and values of their energy spectral
indices converge to $\sigma=2.2$ when $\gamma\rightarrow\infty$ and/or
the magnetic turbulence amplitude grows.

Below, we will present our results on relativistic shock acceleration
published in a series of papers Bednarz \& Ostrowski (1996, 1998, 1999)
and Bednarz (1999). In the next chapter we discuss our numerical
simulations and problems with their application to relativistic shock
conditions. Then, in chapter 3, the acceleration time scales in mildly
relativistic shocks are derived for a number of magnetic field
configurations. Chapter 4 is devoted to ultrarelativistic shocks. We
show convergence of the particle energy spectral index to the asymptotic
value $\sigma_{\infty} \simeq 2.2$ for $\gamma \rightarrow \infty$.
We also discuss particle reflections from large $\gamma$ shocks
providing a limit for models involving GRBs as sources of UHECR.
The acceleration time scale is also derived. In the last chapter 5
a short summary is presented.

\chapter{Numerical simulations}

In order to consider the role of particle anisotropic distributions and
different configurations of the magnetic field in shocks the present
work is based on the small angle particle momentum scattering approach
described by Ostrowski (1991). It enables us to model effects of
cross-field diffusion, important in shocks with oblique magnetic fields.
Let us note (cf. Ostrowski 1993) that this code allows for a reasonable
description of particle transport in the presence of large amplitude
magnetic field perturbations also.

Some earlier derivations of the acceleration time scale were based on
the numerical simulations involving particle scattering at point like
scattering centers isotropizing the particle momentum at each
scattering. This approach does not provide a proper description for the
acceleration processes in shock waves moving with velocities comparable
to the particle velocity because it removes particle anisotropy and
changes factors related to it. Moreover, against arguments presented
in some papers, such scattering pattern can not be realized in turbulent
magnetic fields near relativistic shocks, where most particles active in
the acceleration process are able to diffuse only a short distance below
a few particle gyroradii off the shock\footnote{ However, for the
nonrelativistic shock velocity and particles much above the injection
energy such approximation can be safely used (cf. Jones \&
Ellison 1991).}. Such distances are often insufficient to allow for big
particle pitch-angle changes occurring with the {\em point-like}
scattering centers which isotropize particle momentum at each scattering.
In shocks with oblique magnetic fields such scattering pattern can
substantially change the shape of the accelerated particle spectrum with
respect to the pitch angle diffusion model. Additionally, as an
individual particle interaction with the shock can involve a few
revolutions along the magnetic field, the usually assumed adiabatic
invariant conservation, $p_\perp^2/B = {\rm const}$, cannot be valid for
short inter-scattering intervals.

Below, the light velocity is used as the velocity unit, $c=1$. As the
considered particles are ultrarelativistic ones, $p = E$, we often put
the particle momentum for its energy. In the shock we label all upstream
(downstream) quantities with the subscript `1' (`2'). The quantities are
given in their respective plasma rest frames but subscripts U or D mean
that a parameter is measured in upstream plasma rest frame or
downstream plasma rest frame, respectively.

The shock normal rest frame is the one with the plasma
velocity normal to the shock, both upstream and downstream the shock
(cf. Begelman \& Kirk 1990). The acceleration time scales in relativistic
shocks (chapter 3), $T_{acc}$, are always given in this particular frame
in units of the upstream gyroradius divided by c but the downstream
plasma rest frame quantities are used (chapter 4) for the case of
ultrarelativistic ones, $t_{acc}$.

Here we affix a gyroradius with the index `$g$' when it is a
value given for the local {\em uniform} (tantamount to {\em mean} or
{\em homogeneous}) magnetic field component. Index `$e$' means the
{\em effective} field including the field perturbations (see Eq. 2.15). 

Let us denote parallel diffusion coefficient as $\kappa_{\parallel}$
and perpendicular diffusion coefficient as $\kappa_{\perp}$. Moreover,
we will sometimes use shortcuts $\tau \equiv \kappa_\perp / \kappa_
{\parallel}$ and $\lambda \equiv \log_{10} (\kappa_\perp / \kappa_\| )$.

If it will not cause ambiguity we will use symbol $\psi$ to designate
the magnetic field inclination to the shock normal upstream of the shock,
instead of $\psi_{1}$, and the Lorentz factor of the shock as seen
upstream of the shock as $\gamma$, instead of $\gamma_{1}$.
For the same magnetic field fluctuation patterns upstream and downstream
of the shock we will use symbols without indices for these patterns.

\section{Acceleration time scale in nonrelativistic versus
relativistic shock waves}

In the case of a nonrelativistic shock wave, with velocity $U_1 \ll 1$,
the acceleration time scale can be defined as
$$
T_{acc} \equiv
{E \over {\overline{\Delta E} \over \Delta t}} \qquad , \eqno(2.1)
$$ 
where $\overline{\Delta E}$ is the mean energy gain at particle
interaction with the shock and $\Delta t$ is the mean time between
successive interactions. One can use mean values here because any
substantial increase of particle momentum requires a large number of
shock-particle interactions and the successive interactions are only
very weakly correlated with each other. The respective expression for
$T_{acc}$ in parallel shocks,
$$
T_{acc}^0 = {3 \over U_1-U_2}\, \left\{ {\kappa_1 \over 
U_1} + {\kappa_2 \over U_2} \right\} \qquad , \eqno(2.2)
$$
where $\kappa_i$ is the respective particle spatial diffusion
coefficient, has been discussed by Lagage \& Cesarsky (1983). Ostrowski
(1988) provided the analogous scale for shocks with oblique magnetic
fields and small amplitude magnetic field perturbations. It can be
written in the form
$$
T_{acc}^\psi = {3 \over U_1-U_2} \, \left\{
{ \kappa_{n,1} \over U_1 \sqrt{\kappa_{n,1} \over \kappa_{\parallel,1}
\cos^2{\psi_1}} } +
{ \kappa_{n,2} \over U_2 \sqrt{\kappa_{n,2} \over \kappa_{\parallel,2}
\cos^2{\psi_2}} } \right\}
\qquad , \eqno(2.3)
$$
where the index $n$ denotes quantities normal to the shock, the index
$\parallel$ those parallel to the magnetic field, $\psi$ is an angle
between the magnetic field and the shock normal and $U_1/\cos{\psi_1}
\ll c$ is assumed. The terms $\sqrt{\kappa_n / (\kappa_\parallel
\cos^2\psi)}$ represent a ratio of the mean normal velocity of a
particle to such velocity in the absence of cross-field diffusion. One
may note that for negligible cross-field diffusion the expression (2.3)
coincides with (2.2) if we put $\kappa_{n,i}$ for $\kappa_i$ ($i$ = $1$,
$2$). The case of oblique shock with finite amplitude field
perturbations has not been adequately discussed yet, but we expect the
respective acceleration scale to be between the values given by the
above formulae for $T_{acc}^0$ and $T_{acc}^\psi$. The influence of the
particle escape boundary on the acceleration time scale and the particle
spectrum is discussed by Ostrowski \& Schlickeiser (1996).

\begin{figure}
\vspace*{8.50cm}
\includegraphics{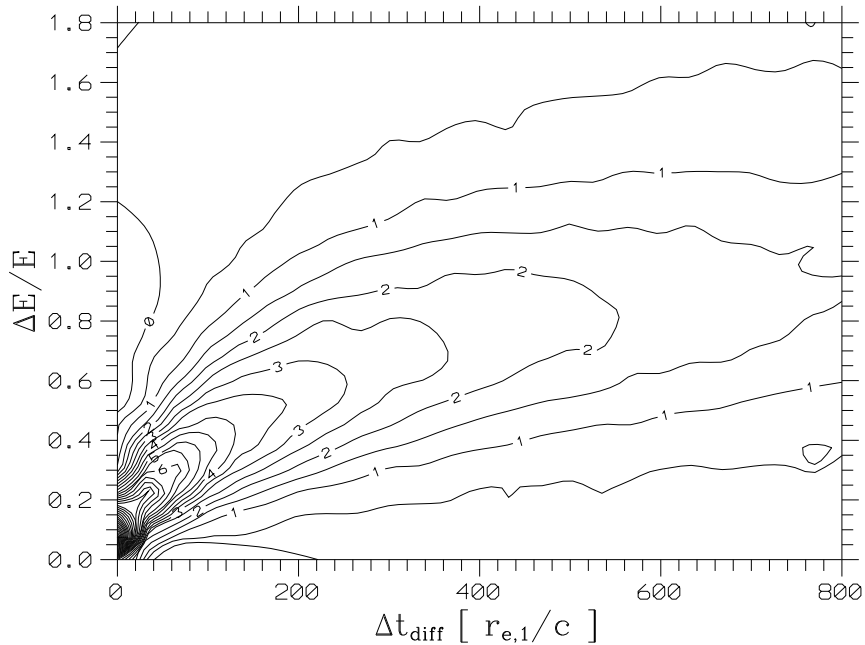}
\caption{Distribution of particle-shock interaction events for upstream
particles, $N(\Delta t_{diff},\Delta E/E)$, in function of the upstream
diffusive times, $\Delta t_{diff}$, and the respective energy changes,
$\Delta E/E$. The correlation between $\Delta t_{diff}$ and $\Delta E/E$
is represented by a regular drift of the distribution maximum toward
higher $\Delta E/E$ when increasing $\Delta t_{diff}$. An example for
the parallel shock with $U_1 = 0.5$, $\psi_1 = 1^\circ$ and weak
scattering conditions ($\kappa_\perp / \kappa_\parallel = 1.6\cdot
10^{-6}$) is presented.}
\label{fig1}
\end{figure}

If the shock velocity becomes relativistic, the particle energy change
at a single interaction with the shock can be comparable, or even larger
than the original energy. Moreover, after interaction with the shock,
the {\em upstream} particles with small initial angles between its
momenta and the mean magnetic field have a larger chance to travel far
away from the shock. On average, such particles spend longer times and
are able to change its pitch angles substantially until the next hits at
the shock. Then, larger pitch angles allow for particle reflections with
large energy gains or for transmissions downstream (cf. Ostrowski 1991,
Lucek \& Bell 1994). Therefore, correlations of the times between
successive interactions, $\Delta t_{diff}$, the energy gains at these
interactions, $\Delta E$, and possibly the probability of particle escape
occur. As an example, in Fig.~2.1 we map the number of particle
interactions with the shock in coordinates ($\Delta t_{diff}$,
$\Delta E$). A cut of the presented surface at any particular value of
$\Delta t_{diff}$ gives the distribution of energy gains for particles
who have spent this time since the last interaction with the shock.
A general trend seen on the map for increasing $\Delta t_{diff}$ is the
growing value of $\Delta E / E$ for the distribution maximum. Because of
these correlations are accompanied with the large energy gains $\Delta
E\sim E$, we propose a different approach to the derivation of the
acceleration time scale with respect to the one used for nonrelativistic
shocks. Usually the acceleration time scale is applied for the
derivation of the highest energies occurring in the particle spectrum,
characterized by its cut-off energy, $E_{c}$. Thus we use this energy
scale to define the acceleration time scale as
$$
T_{acc}^{(c)} \equiv {E_c \over \dot{E}_c } \qquad , \eqno(2.4)
$$
where $\dot{E}_c \equiv dE_c/dt$.  The rate of the cut-off energy
increase is a well-defined quantity and the time scale (2.4) has a clear
physical interpretation. The above definition does not require any limit
for the energy gains of individual particles and all possible
correlations are automatically included here. From the meaning of the
definition (2.4) it follows that $T_{acc}^{(c)}$ is somewhat shorter
than the respective scale at the same energy for later times required
for the respective part of the spectrum to become a pure power-law (cf.
Ostrowski \& Schlickeiser 1996). One should also note that in
relativistic shocks the time scale depends on the reference frame we use
for its measurement. In the present paper the acceleration time scales
are given in the respective normal shock rest frame. However, the
applied time units $r_{e,1}/c$ are defined with the use of
the upstream gyration time.

\section{Transport of particles}

To derive particle trajectories in a disturbed magnetic field one
should, in general, integrate full equation of motion along these
trajectories (see summary in Decker 1988 and Ostrowski 1988).
However, for slightly inhomogeneous fields it was proposed
a `quasi-linear' approximation for analytical calculations
(e.g. Jokipii 1971) consisting of distinguishing between two factors
determining a particle's trajectory: the `adiabatic' undisturbed
motion in the mean field $\vec{B_{0}}$, and perturbations to this
trajectory derived by averaging the effect of magnetic field
perturbations $\delta\vec{B}=\vec{B}-\vec{B_{0}}$ along the trajectory.
As a result the description of particle transport in terms of the
Fokker-Planck equation includes the diffusive term in the pitch
angle $\vartheta$, where $\vartheta\equiv\angle(\vec{p}, \vec{B_{0}})$,
which describes trajectory perturbations and all quantities are
averaged over the phase angle along the trajectory $\varphi$. In the
case of efficient particle scattering maintaining the particle
distribution function $f(\vec{r}, p, \vartheta, t)$ is very nearly
isotropic, the equation can be reduced to the spatial diffusion
equation. Concerning the pitch angle diffusion, all information on the
particle scattering process is contained in the pitch angle diffusion
coefficient $D_{\vartheta}=\langle\Delta\vartheta^{2}\rangle/
(2\langle\Delta t\rangle)$ where $\Delta\vartheta$ ($\ll 1$) is the
change of $\vartheta$ during an individual `scattering act', and
$\langle...\rangle$ denotes taking the average (see Chandrasekhar 1943).
Perturbing force acts at the particle trajectory in a continuous way,
and the notion of the scattering act may be introduced by summing up
all changes to the orbit over some time $\Delta t$, long enough for the
corresponding pitch angle changes to be uncorrelated in the successive
scattering acts. In applications, usually a process of diffusion in
parameter $\mu\equiv\cos\vartheta$ with the corresponding diffusion
coefficient, $D_{\mu}=D_{\vartheta}(1-\mu^{2})$ is considered. The
relation of the above Fokker-Planck approach to the general situation
also involving large-angle scattering was discussed by Kirk \& Schneider
(1988). Our numerical approach resembles the one applied by Kirk \&
Schneider (1987b).

\begin{figure}
\vspace*{8.52cm}
\includegraphics{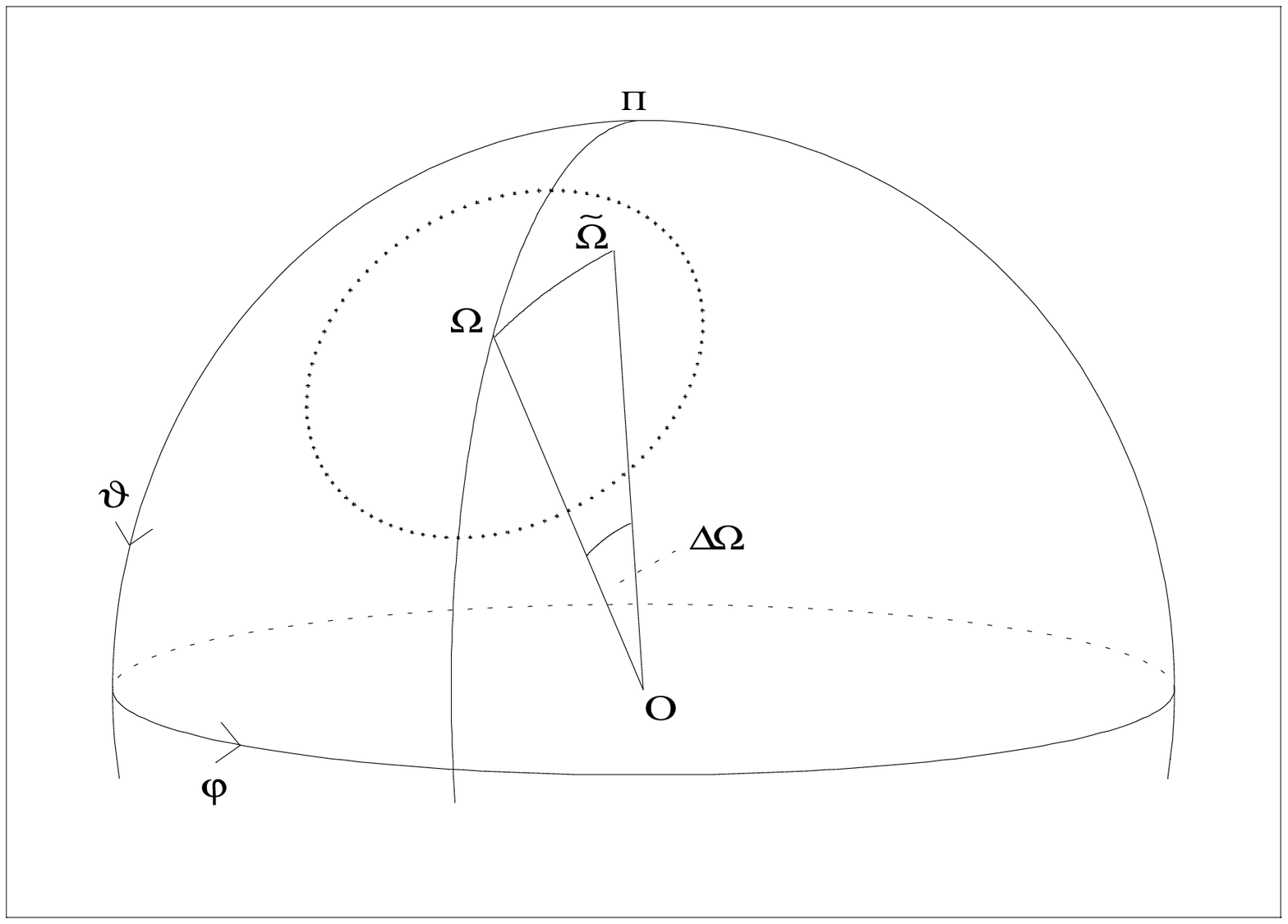}
\caption{
Particle momentum scattering at the sphere $|\vec{p}|$ = const.
The sphere is parameterized with coordinates $\vartheta$ and $\varphi$.
A particle with the original momentum pointing towards $\Omega$ with
coordinates ($\vartheta$, $\varphi$) is scattered to a point
$\tilde{\Omega}$ with coordinates ($\vartheta_{new}$, $\varphi_{new}$),
within a circle at the sphere with radius $\Delta \Omega_{max}$. The
scattering angle is $\Delta\Omega\equiv < (\Omega O \tilde{\Omega})$
and the scattering azimuthal angle
$\beta\equiv < (\Pi\Omega\tilde{\Omega})$.}
\label{fig2o}
\end{figure}
We restrict our consideration to the test-particle approximation, in
which it is assumed that particles are scattered by scattering centers
in the fluid but have no effect either on the fluid velocity or on the
density of scattering centers. Between two successive scatterings, the
particle is assumed to proceed along the undisturbed path in the mean
field. We will furthermore assume that the scattering centers are frozen
into the fluid. The assumption implicates that $|\vec{p}|= {\rm const}$.
We introduce discrete uncorrelated perturbations of the particle's
direction [i.e. perturbations in $\Delta \vartheta$ (or $\Delta \mu$)
and $\Delta \varphi$], in finite time steps, $\Delta t$. Thus, all
particle momentum vectors can be represented as points on the sphere
of constant $|\vec{p}|$ parameterized with two angles $\vartheta$
(or $\mu$) and $\varphi$ (Fig. 2.2). All our calculations were performed
in the respective local plasma rest frame. In any such frame the
electric field vanishes and particle energies are conserved.

If the distribution of the particle orientation $\Omega\equiv
(\vartheta, \varphi)$ at the sphere (Fig. 2.2) maintain the same form
at any point on that sphere independently of the local coordinate
lines then the scattering process is not affected by the orientation.
In the case one gets `isotropic' diffusion coefficient $D_{\vartheta}$
that is independent of $\vartheta$. Let us denote the scattering
amplitude (angle between the original orientation $\Omega$ and the one
after scattering $\tilde{\Omega}$) as $\Delta \Omega$, and the angle
between the meridian, $\varphi = {\rm const }$, and the great circle
connecting $\Omega$ with $\tilde{\Omega}$ as $\beta$. For consistency
we demand that, in the limit of small scattering amplitudes, the
considered scattering model should lead to an isotropic diffusion
coefficient. The considered scattering probability distribution,
$F=F(\Delta\Omega ,\beta)$ must satisfy the `elliptical' symmetry:
$F(\Delta\Omega ,\beta)$=$F(\Delta\Omega ,\pi - \beta)$=$F(\Delta\Omega,
- \beta)$. Kirk \& Schneider (1987b) used the distribution
$F(\Delta\Omega ,\beta)$ derived from the heat conduction equation.
It was equivalent to assuming that the diffusive character of particle
trajectories is also preserved at the limit $\Delta t \rightarrow 0$.
However, in a general case, one should assume a form determined by
considered form of the magnetic field perturbations. The simple choice
is to take $F=F(\Delta\Omega)$ which does not distinguish any direction
in space. Unless one has any particular pattern of field perturbations
it is the most natural choice and we will restrict ourselves to such
distributions below. In particular, we take it in a normalized form
$$F(\Delta\Omega)=
    \left\{
     \begin{array}{ll}
      (1-\cos\Delta\Omega_{max})^{-1}\sin\Delta\Omega &
         \mbox{$\qquad (\Delta\Omega\leq \Delta\Omega_{max})$} \\
      0 & \mbox{$\qquad (\Delta\Omega > \Delta\Omega_{max}$)}
     \end{array}
    \right. \eqno(2.5)
$$
which ensures an equal probability of reaching any unit surface
element of the sphere within the range $\Delta\Omega_{max}$ from the
original position. One should note that this model scattering is no
longer a symmetric one in $\mu$. The fact is visible after averaging
the spherical triangle relation (Eq. 2.11) over $\beta$, for a given
$\Delta\Omega$, the mean change of $\mu$ is $\langle\Delta\mu\rangle=\mu
(\cos\Delta\Omega - 1)$. The anisotropy results from the projection of
the circle $\Delta\Omega = {\rm const}$ in the spherical coordinates
($\mu, \varphi$) and, for constant $\Delta\Omega_{max}$, does not lead
to any actual particle anisotropy. In the Fokker-Planck equation
$$
\frac{\partial f}{\partial t}+v\mu\frac{\partial f}
{\partial z}=\frac{\partial}{\partial \mu} \left (
\frac{\langle\Delta\mu\rangle}{\Delta t} f \right )
+\frac{1}{2}\frac{\partial^2}{\partial \mu^2} \left (
\frac{\langle\Delta\mu^2\rangle}{\Delta t} f \right ) \qquad , \eqno(2.6)
$$
where $f\equiv f(z, \mu, t)$ is the particle distribution function
presented in simplified form with a spatial coordinate $z$ along the
magnetic field, any homogeneous stationary solution must be the isotropic
one. Thus the consistency condition for the Fokker-Planck coefficients is
$$
\langle\Delta\mu\rangle+\frac{1}{2}\frac{\partial}{\partial \mu}
\langle\Delta\mu^2\rangle = 0 \qquad , \eqno(2.7)
$$
and the above mentioned anisotropy is compensated for by a gradient
of the diffusion coefficient $D_{\mu}$.

From the distribution (2.5), for $\Delta\Omega_{max}\ll 1$, one obtains
the relation between mean values $2\langle\Delta\Omega^2\rangle\simeq
4\langle\Delta\vartheta^2\rangle\simeq \Delta\Omega_{max}^2$. Using the
definition $D_{\vartheta}\equiv \langle\Delta\vartheta^2\rangle/(2\langle
\Delta t\rangle)$, we obtain
$$
\Delta\Omega_{max}^2=8D_{\vartheta}\langle\Delta t\rangle \qquad . \eqno(2.8)
$$
A great number of reasonable distributions of $\Delta t$ could be
proposed for any value of $\langle\Delta t\rangle$, which may be
interpreted as representative for different perturbations spectra.
In the limit of infinitesimal scattering amplitude the physical picture
is not sensitive to the particular choice of this distribution. However,
for higher amplitudes and anisotropic particle distributions this
selection may qualitatively affect the simulation results and should be
done with great care (cf. Kirk \& Schneider 1987b, 1988). For the
simulation of large-amplitude scattering one can use equation (2.8) only
in a formal manner. Now, the factor $8D_{\vartheta}$ still provides the
relation between $\Delta \Omega_{max}$ and $\langle\Delta t\rangle$, but
for scatterings of small amplitude it has the additional property of
being 8 times the pitch angle diffusion coefficient. Let us also note
that in our method we make use of the concept of a mean field and assume
that the scatterings are not correlated. For highly perturbed magnetic
fields both assumptions may be of limited validity.

Based on the above model one is able to construct an algorithm for
the derivation of the scattering momentum orientation ($\mu_{new},
\varphi_{new}$) from the original one ($\mu,\varphi$), after an
individual scattering act. Let us denote two independent random values
from the range ($0, 1$) as $R_1$ and $R_2$. Using equation (2.5) we
can generate the value for $\Delta \Omega$ as
$$
\cos(\Delta \Omega)=1-(1-\cos\Delta \Omega_{max})R_1
 \eqno(2.9)
$$
and the orientation angle $\beta$
$$
\beta=2\pi R_2 \qquad . \eqno(2.10)
$$
The new value for pitch angle cosine is derived from the spherical
triangle $\Omega \Pi \tilde{\Omega}$ of Fig 2.2 as
$$
\mu_{new} = \mu \cos\Delta \Omega+\sqrt{1-\mu^2}\sin\Delta \Omega
\cos\beta \qquad . \eqno(2.11)
$$
In equation (2.11) an exact value for $\mu_{new}$ is obtained, and one
can consider high-amplitude scattering ($\Delta \Omega\sim 1$) as well.
However, as was mentioned previously, one should consider carefully the
meaning of the diffusion coefficient in this case. For instance, in
simulating a diffusion perpendicular to the field one should also account
for the possibility of phase perturbation along the trajectory. In our
approach the considered spherical triangle yields
$$
\varphi_{new}=\varphi+\arctan \left (\frac{\sin\Delta \Omega
\sin\beta}{\cos\Delta \Omega\sqrt{1-\mu^2}-\sin\Delta \Omega\mu
\cos\beta} \right ) + \pi H(\mu\mu_{new}-\cos\Delta \Omega)
 \qquad , \eqno(2.12)
$$
where $H(x)$ is the Heaviside step function.

\section{Magnetic field}

In the present discussion we consider the role of the mean magnetic
field configuration and the amount of particle scattering. In order to
avoid effects of varying shock compression due to the presence of
different  magnetic field configurations we take the field as a trace
one without any dynamical effects on the plasma flow. The shock
compression, as seen in the shock normal rest frame, $r=U_{1}/U_{2}$,
is derived from the approximate formulae presented by Heavens \& Drury
(1988). For illustration of the results, in the present theses we
consider the shock waves propagating in the cold electron-proton plasma.
For the {\em mean} magnetic field $B_1$ taken in the upstream plasma
rest frame and inclined at the angle $\psi_1$ with respect to the shock
normal we derive its downstream value and inclination, $B_2$ and
$\psi_2$, with the use of jump conditions presented for relativistic
shocks by e.g. Appl \& Camenzind (1988)
$$
B_2 = B_1 \, \sqrt{ \cos^2{\psi_1} + R^2 \sin^2{\psi_1}} 
\qquad , \eqno(2.13)
$$
$$
\tan{\psi_2} = R \, \tan{\psi_1}  \qquad , \eqno(2.14)
$$
where $R = r \, \gamma_1 / \gamma_2$ and the Lorentz factors $\gamma_i
\equiv 1/\sqrt{1-U_i^2}$ ($i$ = $1$, $2$). These formulae are valid for
both sub- and super-luminal magnetic field configurations.

We model particle trajectory perturbations by introducing small-angle
random momentum scattering along the mean-field trajectory (cf. Ostrowski
1991). The particle momentum scattering distribution is uniform within
a cone wide at $\Delta \Omega$, along the original
momentum direction. The presented simulations for mildly relativistic
shocks use a constant value of $\Delta \Omega = 0.173$ ($ = 10^\circ$).
Scattering events are at discrete instants, equally spaced in time as
measured in the units of the respective $r_{g,i}/c$ ($i$ = $1$, $2$).
The increasing perturbation amplitude is reproduced in simulations by
decreasing the time period $\Delta t$ between the successive scatterings.

In ultrarelativistic shock waves efficient particle scattering with a
very small $\Delta \Omega$ requires derivation of a large number of
scattering acts and the respective numerical code becomes extremely
time-consuming. In order to overcome this difficulty we propose a hybrid
approach involving `very small' $\Delta \Omega_{C}$ ($\sim 0.5\gamma^{-1}$)
close to the shock, where the scattering details play a role, and much
larger scattering amplitude $\Delta \Omega_{F} = 9^\circ$ to describe
particle diffusion further away from the shock. The respective
scaling of the scattering time $\Delta t$ is performed in both cases
($\Delta \Omega_{C}^2/\Delta t_{C}=\Delta \Omega_{F}^2/\Delta t_{F}$) to
yield the same turbulence amplitudes measured by the values of the
cross-field diffusion coefficient, $\kappa_\perp$ and the parallel
diffusion coefficient $\kappa_\|$. For a few instances we checked the
validity of this approach by reproducing the results for the small
$\Delta \Omega_{C}$ everywhere.

For simplicity, except sections 4.1 and 4.3, we use the same scattering
pattern ($\Delta \Omega$ and $\Delta t$ in units of $r_g/c$) upstream
and downstream the shock, leading to the same values of $\kappa_\perp /
\kappa_\parallel$ in these regions (see, however, Ostrowski 1993). One
should note that the particle momentum scattering due to the presence of
the turbulent magnetic field is equivalent to the effective magnetic
field larger than the respective uniform mean component, $B_1$ or $B_2$.
In our model, the effective field can be estimated as
$$
B_{e,i} = B_i \sqrt{1 + \left( 0.67 {\Delta \Omega \over \Delta t}
\right)^2} \quad (i = 1, 2) \qquad . \eqno(2.15)
$$
It is the lower limit for the actual field since the amount of power in
perturbations with wave-lengths smaller than $c \, \Delta t$ cannot be
considered within such a simple model. The amount of energy in magnetic
turbulence with the waves shorter than $c\, \Delta t$ is required to be
small because the presented estimate assumes the particle momentum
perturbation in $\Delta t$ occurs on the uniform effective perturbing
field. To compare the scattering processes with different $\Delta t$ one
has to neglect the unknown factor of the ratio of the averaged actual
magnetic field to the estimated value (like the one in Eq.~2.15). Let us
note that this factor, as well as the notion of the effective field were
not considered earlier.

For relativistic shocks the derived acceleration time scales are
presented in units of the formal diffusive scale $T_0 \equiv
4( \kappa_{n,1} / U_1 + \kappa_{n,2} / U_2 ) / c$ or in units of
$r_{e,1}/c$ , in the shock normal rest frame but for ultrarelativistic
ones in units of $r_{g,2}/c$ in the downstream plasma rest frame.

\section{Fitting\hspace{0.2em} the\hspace{0.2em} spectrum\hspace{0.2em}
and\hspace{0.2em} the\hspace{0.2em} acceleration\hspace{0.2em} time}

Our numerical calculations involve particles with momenta systematically
increasing over several orders of magnitude. In order to avoid any
energy dependent systematic effect we consider the situation with all
spatial and time scales -- defined by the diffusion coefficient, the
mean time between scatterings and the shock velocity -- to be proportional
to the particle gyroradius, $r_g = p/(eB)$, i.e. to its momentum.

\begin{figure}
\vspace*{7.02cm}
\includegraphics{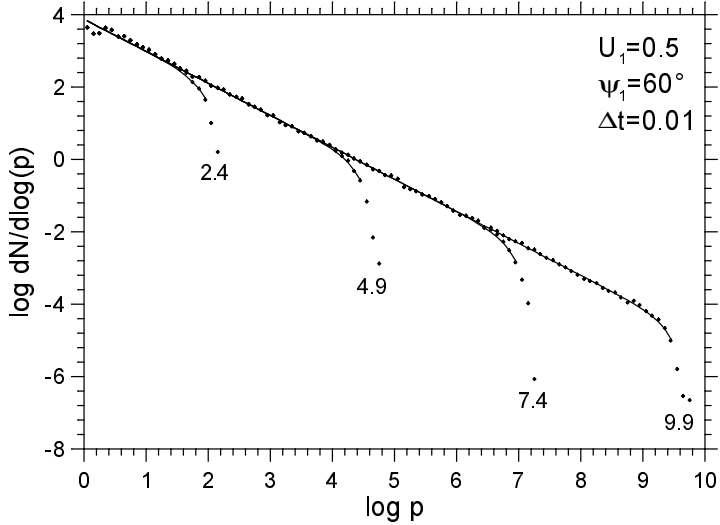}
\caption{Particle spectra at a sequence of time instants. 
Individual simulation points represent the particle number (weight)
$\Delta N$ per logarithmic momentum bin $\Delta \log p$. With full 
lines we present the respective fits (2.16).}
\label{fig3}
\end{figure}

For a chosen shock velocity and the magnetic field configuration we
inject particles in the shock at some initial momentum $p_0$ and follow
their phase-space trajectories. We assume the constant particle
injection to continue in time after the initial time $t_0 = 0$.
A particle is excluded from simulations if it escapes through the 
free-escape boundary placed far downstream of the shock or reaches the
energy larger than the assumed upper limit. These particles are replaced
with the ones arising from splitting the remaining high-weight particles,
preserving their physical parameters (cf. Kirk \& Schneider 1987b;
Ostrowski 1991). Particles that exist longer than the time upper limit
for simulations are excluded from simulations without replacing. Here we
put the boundary at the distance $6\kappa_{2,n}/U_2 + 4 r_{g,2}$ for
relativistic and $4 r_{g,2}$ for ultrarelativistic shocks. We checked by
simulations that any further increase of this distance does not
influence the results in any noticeable way. For every shock crossing,
the particle weight factor multiplied by the inverse of the particle
velocity normal to the shock ($\equiv$ particle density) is added to the
respective time and momentum bin of the spectrum as measured in the
shock normal rest frame. As one considers a continuous injection in all
instants after $t_0$, in order to obtain the particle spectrum at some
time $t_j > t_0$ one has to add to particle density in a bin $p_i$ at
$t_j$ the densities in this momentum bin for all the earlier times. The
resulting particle spectra are represented as power-law functions with
the squared exponential cut-off in momentum
$$
f(p,t) = A \, p^{-\alpha} \, e^{-\left( p \over p_c\right)^2}
\qquad . \eqno(2.16)
$$
In this formula three parameters are to be fitted: the normalization
constant $A$, the spectral index for the stationary solution $\alpha$,
and the momentum cut-off $p_c$ (Fig.~2.3). Any simulated spectrum
evolved in time by increasing the width of its power-law section and
thus the best fit of this power-law was possible with the use of the
final spectrum at maximum time. Therefore, in the simulations we used
the last spectrum to fit parameters $A$ and $\alpha$. Next, for any
earlier spectrum, these parameters were assumed to be constant and we
were fitting only the cut-off momentum, $p_c$. For each fit we used 20
last points of the spectrum preceding the point where particle density
fell below $0.16$ of $A\,p^{-\alpha}$. The number of $0.16$ was chosen
experimentally in order to obtain the best fits to the cut-off region
of the spectrum. As the distribution (2.16) represents only an
approximation to the actual particle distribution, there was no reason
to use points corresponding to lower densities of lesser statistical
significance.

\begin{figure}
\vspace*{7.0cm}
\includegraphics{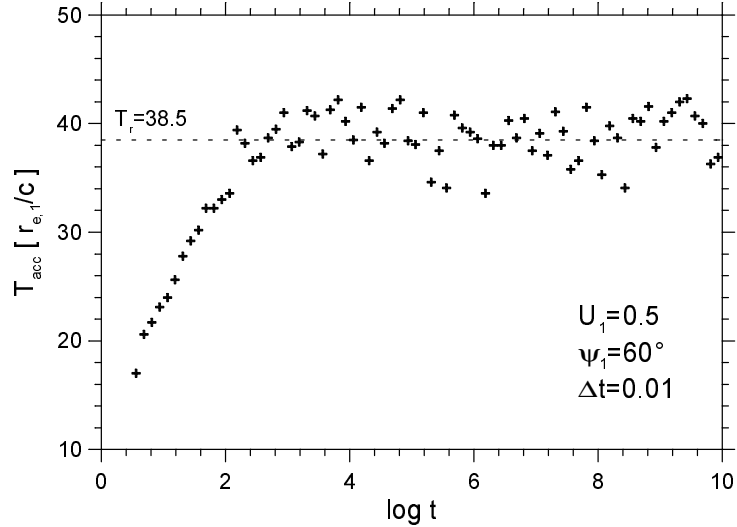}
\caption{An example of simulated values of $T_{acc}^{(c)}$ in units of
$r_{e,1}/c$ in a single run of the code. The simulation time $t$ is
given in units of $r_{g,1}(p=p_0)/c$ in the shock normal rest frame. We
fit the final acceleration time scale only to the points at the advanced
phase of acceleration (dashed line). The dispersion of these points defines
the fitting error.}
\label{fig4}
\end{figure}

In the simulations, due to our proportional momentum scaling of the
respective quantities, the derived acceleration time scale (2.4) must be
also proportional to $p$, and thus to $r_g(p)/c$. Therefore, this time
scale measured in units of $r_g(p_c)/c$ (or $r_e(p_c) /c$) is momentum
independent and can be easily scaled to any momentum. The parameter
$T_r$ gives the value of the acceleration time scale in units of
$r_{e,1}/c$, i.e. $T_{acc}^{(c)} = T_r \, r_{e,1}/c$. The value of
$T_{acc,i}^{(c)}$ at a particular time $t_i$ is derived from the
respective values of $p_{c,i}$:
$$
T_{acc,i}^{(c)} =
{p_{c,i} \over {p_{c,i}-p_{c,i-1} \over t_i-t_{i-1} }}
\qquad , \eqno(2.17)
$$
where we consider the advanced phase of acceleration ($p_{c,i} \gg
p_0$). As in our simulations $p_c \propto t$ the condition
$(p_{c,i}-p_{c,i-1})/p_{c,i} \ll 1$ is not required to hold in equation
(2.17). Therefore, with all scales proportional to the particle momentum,
the formula (2.17) reduces to $T_{acc,i}^{(c)} = t_i$ and the parameter
$T_r$ tends to a constant (Fig.~2.4). The extension of the simulated
spectra over several decades in particle energy allows to avoid problems
with the initial conditions and decrease the relative error of the
derived time scale by averaging over a larger number of instantaneous
$T_{acc,i}$.

\chapter{The acceleration time scale in relativistic shock waves}

For a given relativistic shock velocity  particle anisotropy in the
shock depends on the mean magnetic field inclination to the shock normal
and the form of turbulent field. Below, we describe the results of
simulations performed in order to understand the time dependence of the
acceleration process in various conditions. In order to do that we
consider shock waves propagating with velocities $U_1$ = $0.3$, $0.5$,
$0.7$ and $0.9$ of the velocity of light and the magnetic field
inclination: $\psi_1$ = $1^\circ$, $25.8^\circ $, $45.6^\circ $,
$60^\circ $, $72.5^\circ $, $84.3^\circ$ and $89^\circ$. The first
one is for a parallel shock, the last two ones are for perpendicular
super-luminal shock with all velocities $U_1$. The intermediate values
define luminal shocks ($U_1 / \cos \psi_1 = 1.0$) at the successive
velocities considered, respectively $U_1$ = $0.9$, $0.7$, $0.5$ and $0.3$.

In all these cases we investigate the role of varying magnitude of
turbulence characterized here by the value of $\Delta t$ or by the
ratio of the diffusion coefficient across the mean field and that along
the field, $\kappa_\perp / \kappa_\parallel$ . The relation between
these parameters for $\Delta \Omega = 10^\circ$ is presented in
Fig.~3.1 where at $\Delta t > 0.01$ the presented relation has the
power-law form $\kappa_\perp / \kappa_\parallel = 6.3 \cdot 10^{-5} \,
(\Delta t)^2$.

\section{Parallel shocks}

The most simple case for discussion of the first-order Fermi
acceleration is a shock wave with parallel configuration of the mean
magnetic field. As an example we consider the shock with negligible
field inclination $\psi_1 = 1^\circ$. For such a shock, the present
simulations confirm the expected relation of decreasing the
acceleration time scale with increasing the shock velocity and the
amplitude of trajectory perturbations (Fig.~3.2). One should note at the
upper panel of the figure that for short $\Delta t$ the presented time
scales decrease more and more slowly with decreasing $\Delta t$. It is
due to the fact that starting from some value of $\Delta t$ we reach
conditions of nearly isotropic diffusion, $\kappa_\parallel \approx
\kappa_\perp$ and further decreasing of the time delay between
scatterings decreases the acceleration time in much the same proportion
as the time unit $r_{e,1}/c$ used to measure it (cf. Eq.~2.15, Fig.~3.1).
In the lower panel of Fig.~3.2 the diffusive time scale $T_0$
($\equiv 4( \kappa_{n,1} / U_1 + \kappa_{n,2} / U_2 ) / c$) is used as
the time unit. The minute differences between the successive curves
reflect the statistical fluctuations arising during simulations. Without
such fluctuations all curves should coincide. The one sigma fit errors
of $T_{acc}^{(c)}$ are indicated near the respective points. One should
note that for increasing the shock velocity the acceleration time scale
decreases with respect to the diffusive time scale.

\begin{figure}
\vspace*{7.0cm}
\includegraphics{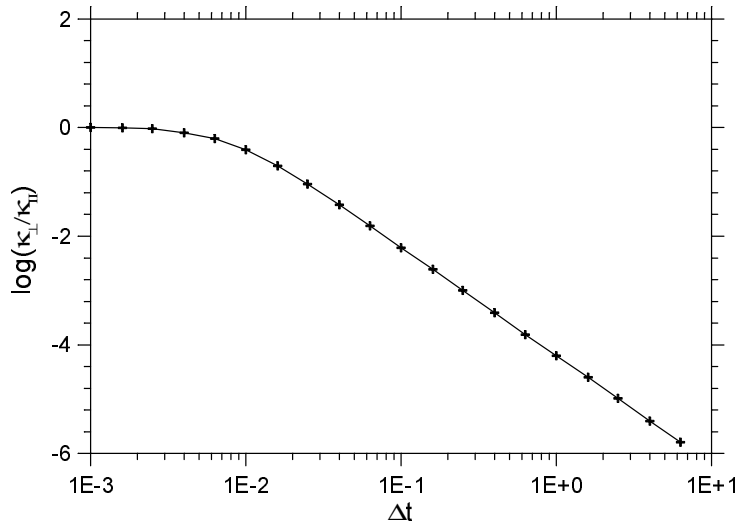}
\caption{The relation between the scattering parameter $\Delta t$ and 
the respective ratio of diffusion coefficients $\kappa_\perp / 
\kappa_\parallel$.}
\label{fig5}
\end{figure}

\section{Variation of $T_{acc}^{(c)}$ with magnetic field
inclination}

In order to compare the acceleration time scales for different magnetic
field inclinations $\psi_1$ we performed simulations assuming a constant
scattering parameters upstream and downstream yielding the same ratio
of $\kappa_\perp / \kappa_\parallel$ in these regions. However, due to
shock compression the particle gyration period is shorter downstream
than upstream in proportion to the {\em mean} magnetic field compression
(Eq.~2.13). In Fig.~3.3 we present the values of the acceleration time
scale derived in such conditions at different $\psi_1$. For
super-luminal shocks the results are presented for the cases allowing
for particle power-law energy spectra, i.e. when the cross-field
diffusion is sufficiently effective. Actually, the spectra with
inclinations $\alpha < 10.0$ are only included.

In general, the acceleration time scale decreases with increasing field
inclination, reaching in some cases the values comparable, or even
smaller than the particle upstream gyroperiod (6.28 in our units of
$r_{e,1}/c$). The trend can be reversed for intermediate wave amplitudes
when the magnetic field configuration changes into the luminal and
super-luminal one. Such changes are accompanied with the steepening of
the spectrum (see below). The acceleration rate at different scattering
amplitudes changes with $\psi_1$ in a way that at different inclinations
the minimum acceleration times occur at different perturbation
amplitudes (different $\Delta t$).

\begin{figure}
\vspace*{11.6cm}
\includegraphics{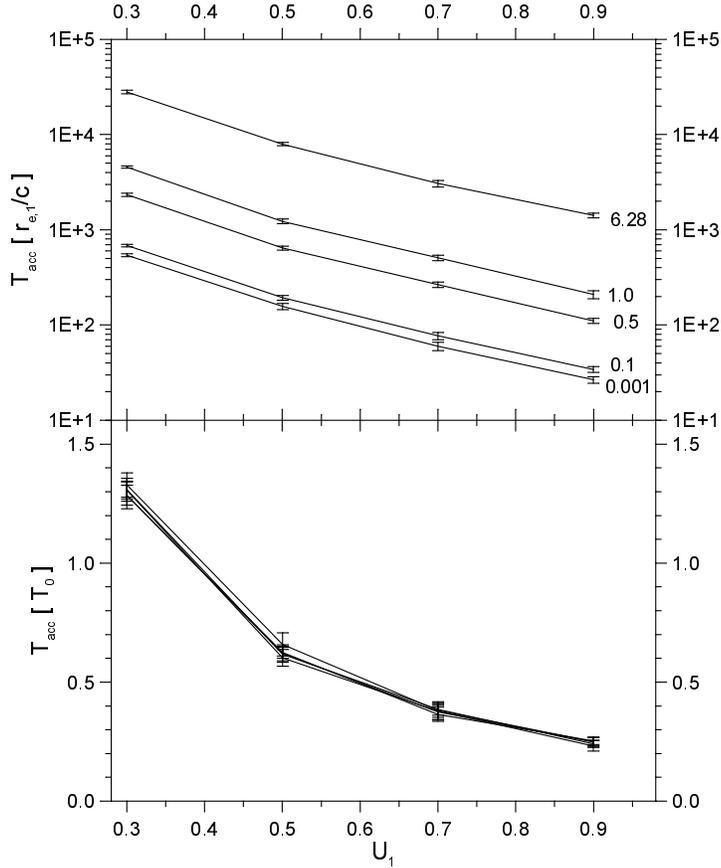}
\caption{The values of $T_{acc}^{(c)}$ in parallel shock waves ($\psi_1
= 1^\circ$) in units of a.) $r_{e,1} / c$ and b.) $T_0$ versus the shock
velocity $U_1$. The values resulting from simulations are given for $U_1
= 0.3$, $0.5$, $0.7$ and $0.9$ with the scattering amplitude parameter
$\Delta t$ near the respective results at the upper panel.}
\label{fig6}
\end{figure}

An important feature  of the acceleration process in relativistic shocks
should be mentioned at this point. The variations of $T_{acc}^{(c)}$ in
oblique shocks are accompanied by changes of the particle spectrum
inclination (cf. Kirk \& Heavens 1989; Ostrowski 1991). In Fig.~3.4, the
curves at ($T_{acc}^{(c)}$, $\alpha$) plane represent the results for
decreasing the scattering amplitude expressed with parameter $\Delta t$,
and joined with lines for the same magnetic field inclination $\psi_1$ .
For parallel shocks the changes in $T_{acc}^{(c)}$ do not lead to any
variation of the spectral index. However, for oblique sub-luminal
($\psi_1$ = $25.8^\circ$, $45.6^\circ$) and luminal ($\psi_1$ =
$60^\circ$) shocks a non-monotonic behavior is seen. The trend in
changing $T_{acc}^{(c)}$ and $\alpha$ observed at smaller perturbation
amplitudes (larger $\Delta t$) is reversed at larger amplitudes when
the substantial cross-field diffusion is possible.

\begin{figure}[t]
\vspace*{10.0cm}
\includegraphics{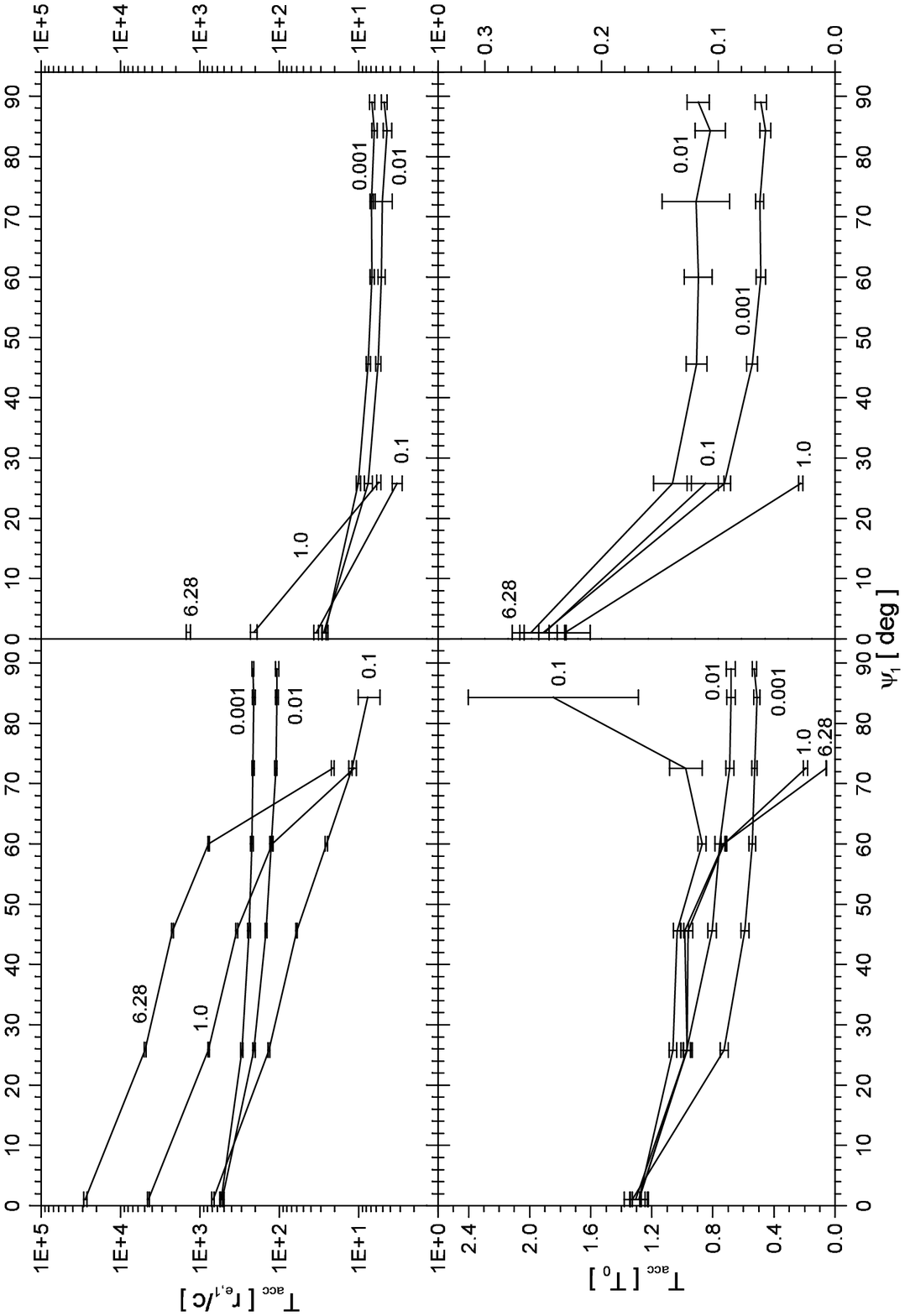}
\caption{The value of $T_{acc}^{(c)}$ in units of $r_{e,1} / c$ (upper
panels) and $T_0$ (lower panels) versus the magnetic field inclination
$\psi_1$. The values resulting from simulations are presented for $U_1 =
0.3$ (left panels) and $0.9$ (right panels) with values of the parameter
$\Delta t$ given near the respective results.}
\label{fig7}
\end{figure}

For oblique shocks (cf. Fig.~3.5) we observe an analogous reduction of
the acceleration time scale as that reported by Naito \& Takahara (1995)
with the pitch angle diffusion model allowing for a more rapid
acceleration than the large angle scattering model. Of course this
agreement is broken for short $\Delta t$, where the cross field
diffusion can not be neglected and the particle magnetic momentum is not
conserved at interactions with the shock.

\section{Variation of $T_{acc}^{(c)}$ with varying turbulence levels}

In a parallel shock the acceleration time scale reduces with the
increased turbulence level in it's neighborhood. This phenomenon, well
known for nonrelativistic shocks (cf. Lagage \& Cesarsky 1983), is
confirmed here for relativistic shock velocities (Fig.~3.2). In general,
there are two main reasons for this change. The first one is a simple
reduction of the diffusion time of particles outside the shock due to
shorter intervals between scatterings analogous to the decrease
observed in nonrelativistic shocks. However, the increased amount of
scattering influences also the acceleration process due to changing
(decreasing) the particle anisotropy at the shock and thus,
\begin{figure}[t]
\vspace*{7.05cm}
\includegraphics{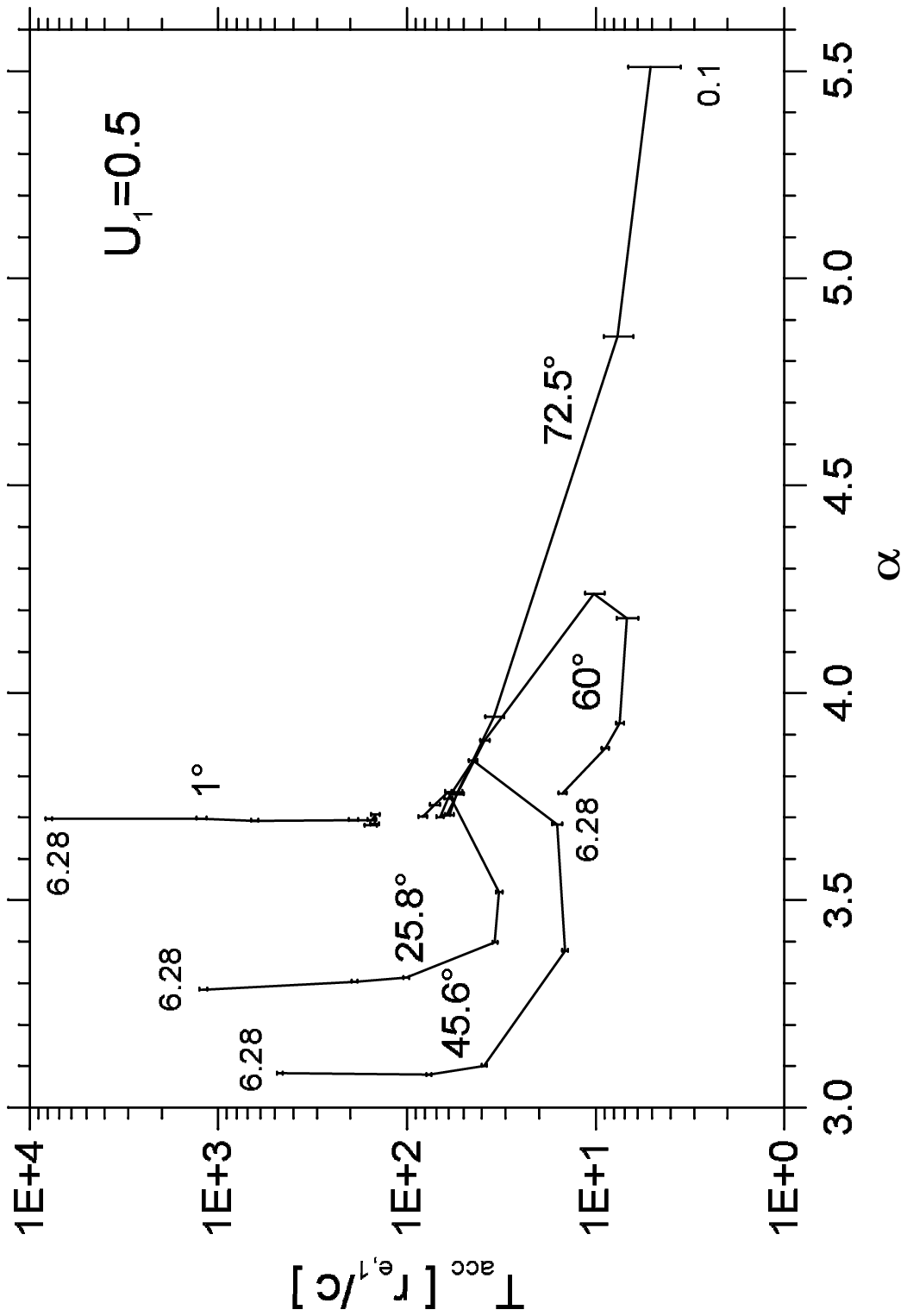}
\caption{The values of $T_{acc}^{(c)}$ in units of $r_{e,1} / c$ at
different inclinations $\psi_1$ versus the particle spectral index
$\alpha$. The values resulting from simulations are given for $U_1$ =
$0.5$ for five values of the angle $\psi_1$ given near the respective
results. The {\em maximum} value of $\Delta t$ is given at the end of
each curve and it monotonously decreases along the curve. }
\label{fig8}
\end{figure}
modifying the mean energy gain of particles interacting with the shock
discontinuity. Additionally, in oblique shocks the upstream-downstream
transmission probability may increase. One should note that the present
approach is not able to describe fully the effect of decreasing
anisotropy with the small amplitude random scattering model applied. It
is due to the fact that correlations between the successive
modifications of a trajectory (a sequence of small angle scattering acts
in this paper) in a single MHD wave cannot be accurately modeled within
the simplified approach used. A more exact approach requires integration
along the particle trajectories in realistic configurations of the
magnetic field. However, the comparison of the present simplified method
to the one involving such an integration shows a reasonably good
agreement (Ostrowski 1993) suggesting that averaging over realistic
trajectories is equivalent in some way to such averaging within our
random scattering approach.

In shocks with oblique magnetic fields a non-monotonic change of the
acceleration time scale with the amount of scattering along the particle
trajectory is observed (Fig.~3.6, see also Fig.~3.3; cf. Ostrowski 1991
for the spectral index). Increasing the amount of turbulence up to
some critical amplitude decreases the diffusion time along the magnetic
field and thus $T_{acc}^{(c)}$. However, as the mean diffusion time
outside the shock is related to the normal diffusion
coefficient\footnote{One should note that for the relativistic shocks,
due to particle anisotropy, the respective relation may be not so simple
as that given in equation (2.2) for nonrelativistic shocks.} $\kappa_n$
($\kappa_{n,i} = \kappa_{\parallel,i} \cos^2{\psi_i} + \kappa_{\perp,i}
\sin^2{\psi_i}$, $i$ $=$ $1$, $2$), the increasing $\kappa_\perp$ will
lead,\pagebreak[4] for large scattering amplitudes to
\begin{figure}[t]
\vspace*{7.0cm}
\includegraphics{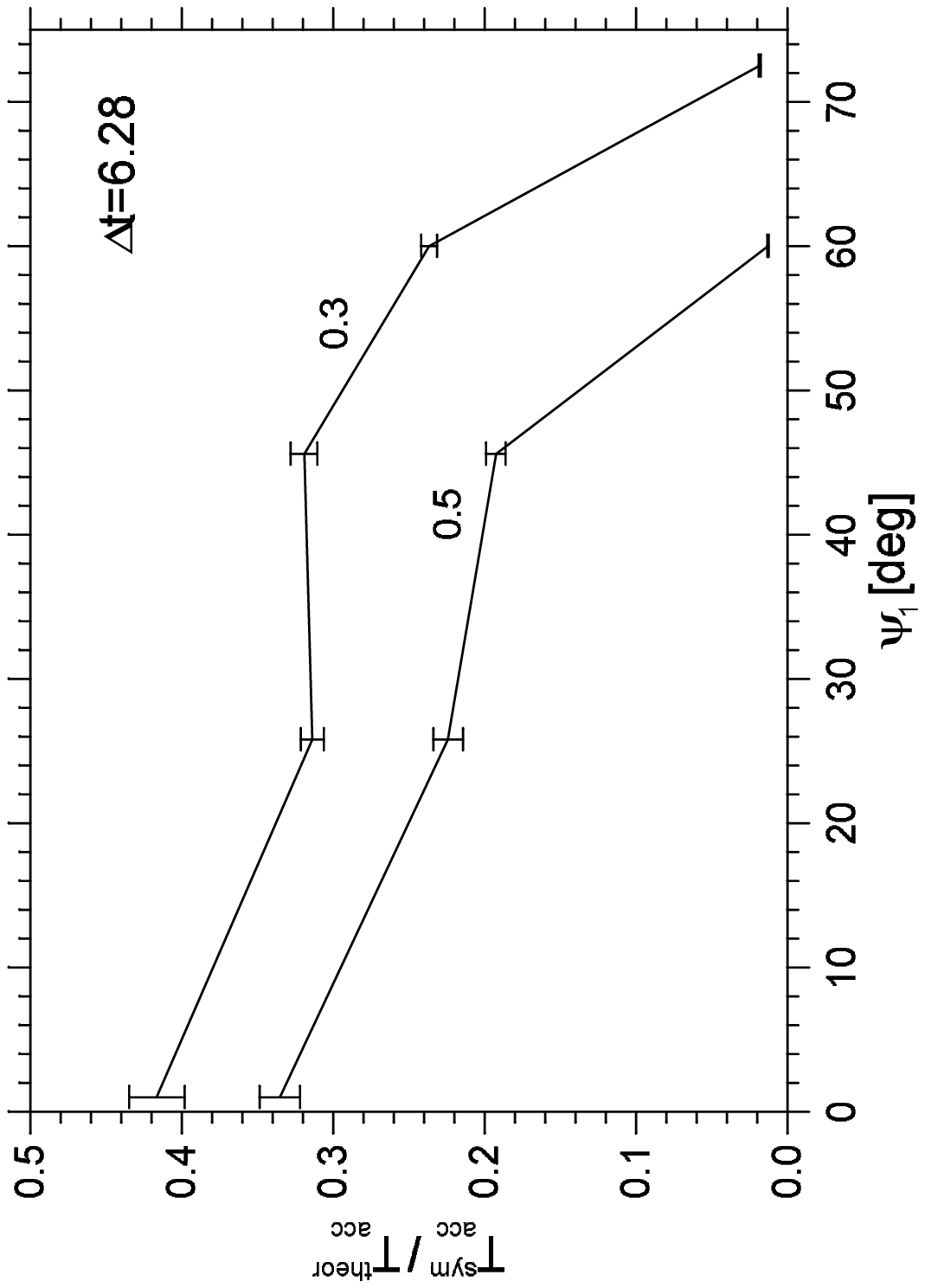}
\caption{The ratio of the pitch angle diffusion simulated time scale
$T_{acc}^{sym}$ to the `theoretical' nonrelativistic time scale
$T_{acc}^{theor}$ given in Eq.~2.2 for different inclinations of the
magnetic field $\psi_1$. The results for $U_1 = 0.3$ and $0.5$ are
presented for the case of negligible cross-field diffusion ($\Delta t
= 6.28$ in the pitch angle diffusion model).}
\label{fig10}
\end{figure}
\begin{figure}
\vspace*{7.0cm}
\includegraphics{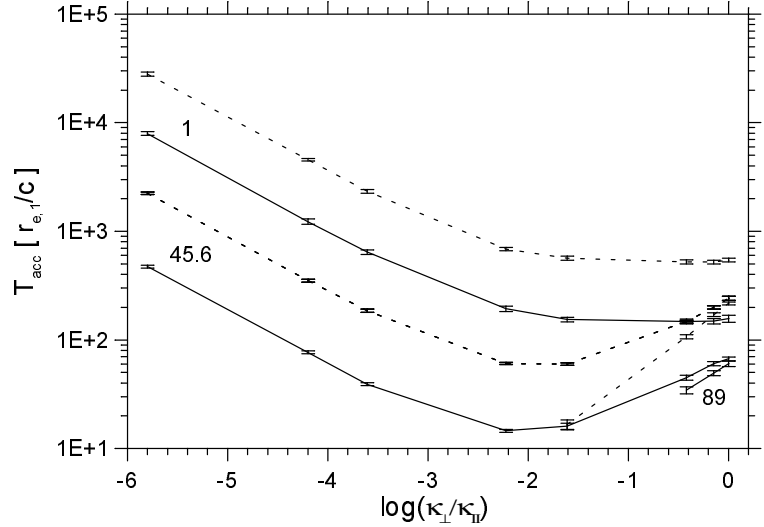}
\caption{The acceleration time $T_{acc}^{(c)}$ versus the level of
particle scattering $\kappa_\perp / \kappa_\parallel$ for shocks with
velocity $U_1$ $ =$ $0.3$ (dashed lines) and $0.5$ (full lines). We
present results for three values of the magnetic field inclination: a.)
parallel shock ($\psi_1 = 1^\circ$), b.) a sub-luminal shock with
$\psi_1 = 45.6^\circ $ and c.) a super-luminal shock with $\psi_1 =
89^\circ $.}
\label{fig9}
\end{figure}
\clearpage
\noindent
longer $T_{acc}$ in units
of $r_e/c$. In the units of $T_0$ the acceleration time depends only
weakly on the turbulence level and shows a small maximum for the minimum
at the presented figure. For super-luminal shocks one can note the
absence of data points corresponding to low turbulence levels, where the
power-law spectrum cannot be formed or it is extremely steep. In these
excluded cases, the upstream population of energetic particles is only
compressed at the shock with the characteristic upstream time of $\sim
r_{e,1}/U_1$ (cf. Begelman \& Kirk 1990; Ostrowski 1993).

\chapter{Ultrarelativistic shock waves}

In the present chapter we discussed several aspects of the first order
acceleration process active at ultrarelativistic ($\gamma\gg 1$) shock
waves. These results are partly published in Bednarz \& Ostrowski (1998,
1999) and in Bednarz (1999).

Below the downstream magnetic field is derived for the relativistic
shock with the compression $R$ obtained with the formulas of Heavens
\& Drury (1988) for a cold ($e$, $p$) plasma -- $R \approx 3.6$ for our
smallest value of $\gamma = 3$ and tends to $R = 3$ for $\gamma >> 1$,
as measured in the shock rest frame.

\section{Acceleration mechanism}

A particle crossing the shock to upstream medium has a momentum vector
nearly parallel to the shock normal. Then the particle momentum changes
its inclination in two ways by: 1) scattering in an inhomogeneous magnetic
field and 2) smooth variation in a homogeneous field component. Hereafter,
the mean deflection angle in these two cases will be denoted by
$\Delta \Omega_{S}$ and $\Delta \Omega_{H}$, respectively. The first
process is a diffusive one and the second depends on time linearly. That
means that with increasing shock velocity, keeping other parameters
constant, $\Delta \Omega_{S}$ decreases slower as a square root of time
in comparison with $\Delta \Omega_{H}$. The Lorentz transformation shows
that with $\gamma\gg 1$ even a tiny angular deviation in the upstream
plasma rest frame can lead to a large angular deviation in the downstream
plasma rest frame. Let us denote a particle phase by $\varphi$ and the
angle between momentum and a magnetic field vector by $\vartheta$ both
measured in the downstream plasma rest frame. Values of these parameters
at the moment when a particle crosses the shock downstream determine if it
is able to reach the shock again in the case of neglected magnetic field
fluctuations downstream of the shock. In fact a motion in the homogeneous
magnetic field carries a particle in such a way that in most cases it
cannot reach the shock again. The magnetic field fluctuations perturbing
the momentum direction lead to broadening the ($\varphi,\vartheta$) range
that allows particles to reach the shock again. Thus, as we show below
for efficient scattering, when $\Delta \Omega_{H}$ becomes unimportant in
comparison to $\Delta \Omega_{S}$, the spectral index and the acceleration
time reach their asymptotic values. The discussed relation between
$\Delta \Omega_{H}$ and $\Delta \Omega_{S}$ is reproduced in our
simulations and presented in Fig. 4.1. There are shown 11 points from
$\gamma$ = 100 to 320 and three other for $\gamma$ = 640, 1280, 2560. The
expected linear dependence of these quantities can be noticed.

\begin{figure}
\vspace*{7.0cm}
\includegraphics{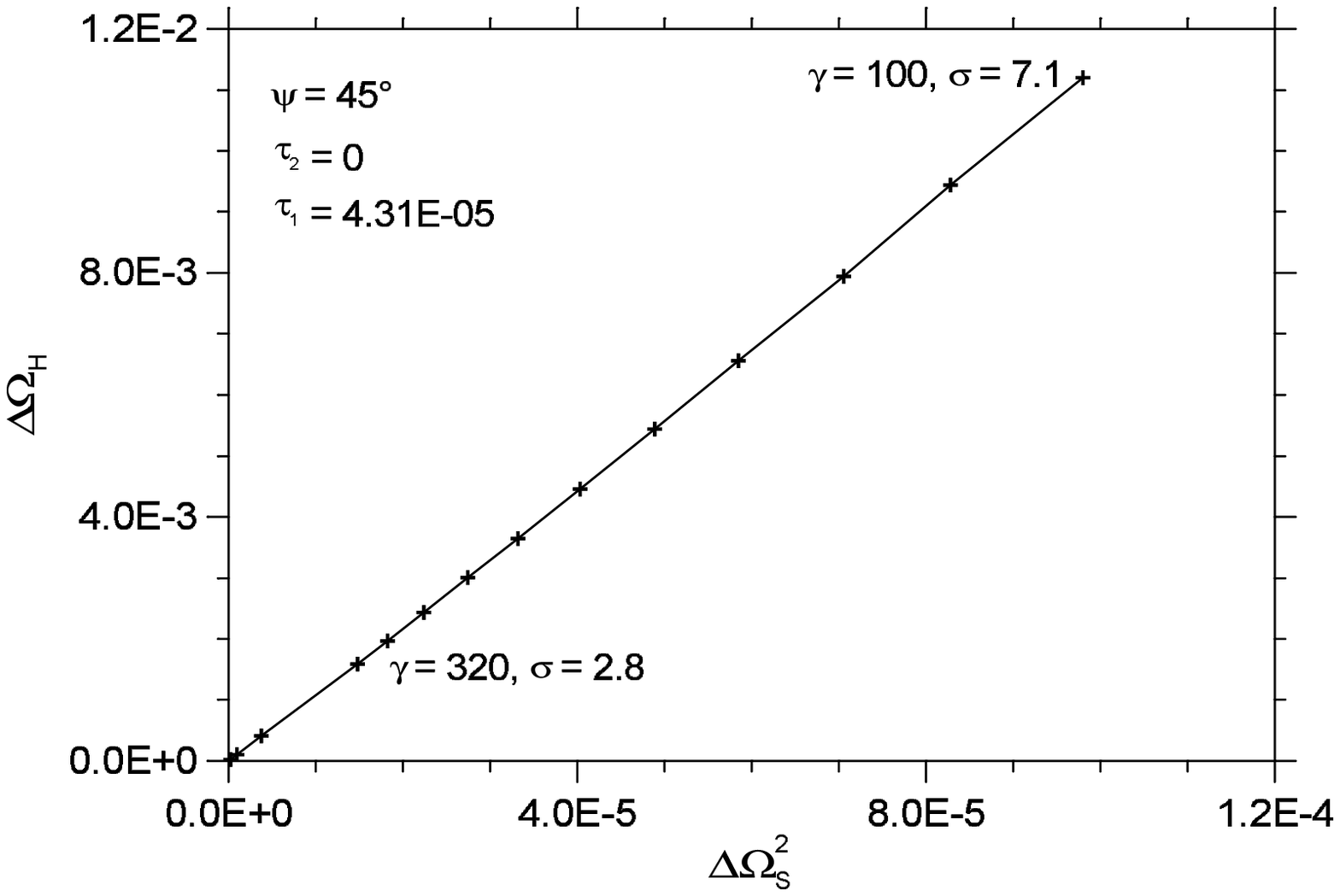}
\caption{The relation between the mean deflection angle upstream of
the shock caused by the scattering in an inhomogeneous
magnetic field ($\Delta \Omega_{S}$) and by smooth variation
in a homogeneous magnetic field ($\Delta \Omega_{H}$).
Last three points for $\Delta \Omega_{H}$ below $1\cdot 10^{-3}$
represent $\gamma$ = 640, 1280, 2560 and yield $\sigma$ =
2.5, 2.3 and 2.2 respectively.}
\label{fig11}
\end{figure}

\section{Energy spectra}

Particle spectral indices were derived for different mean magnetic
field configurations, measured by the magnetic field inclination $\psi$
with respect to the shock normal in the upstream plasma rest frame and
for different amounts of turbulence measured by $\kappa_\perp /
\kappa_\|$. In the simulations we considered a few configurations of
the upstream magnetic field with inclinations with respect to the shock
normal being $\psi$ = $0^\circ$, $10^\circ$, $20^\circ$, $30^\circ$, 
$60^\circ$ and $90^\circ$. The first case represents the parallel shock,
the second is for the oblique shock - subluminal at $\gamma = 3$ and a
superluminal one at larger $\gamma$, and the larger $\psi$ are for
superluminal perpendicular shocks for all considered velocities. We
applied the same patterns upstream and downstream of the shock for the
fluctuation levels
$\lambda = -5.34, -4.39, -3.44, -2.49, -1.56, -0.67, -0.16, 0.00$
($\lambda\equiv\log_{10} \kappa_\perp / \kappa_\| $).

\begin{figure}
\vspace*{14.5cm}
\includegraphics{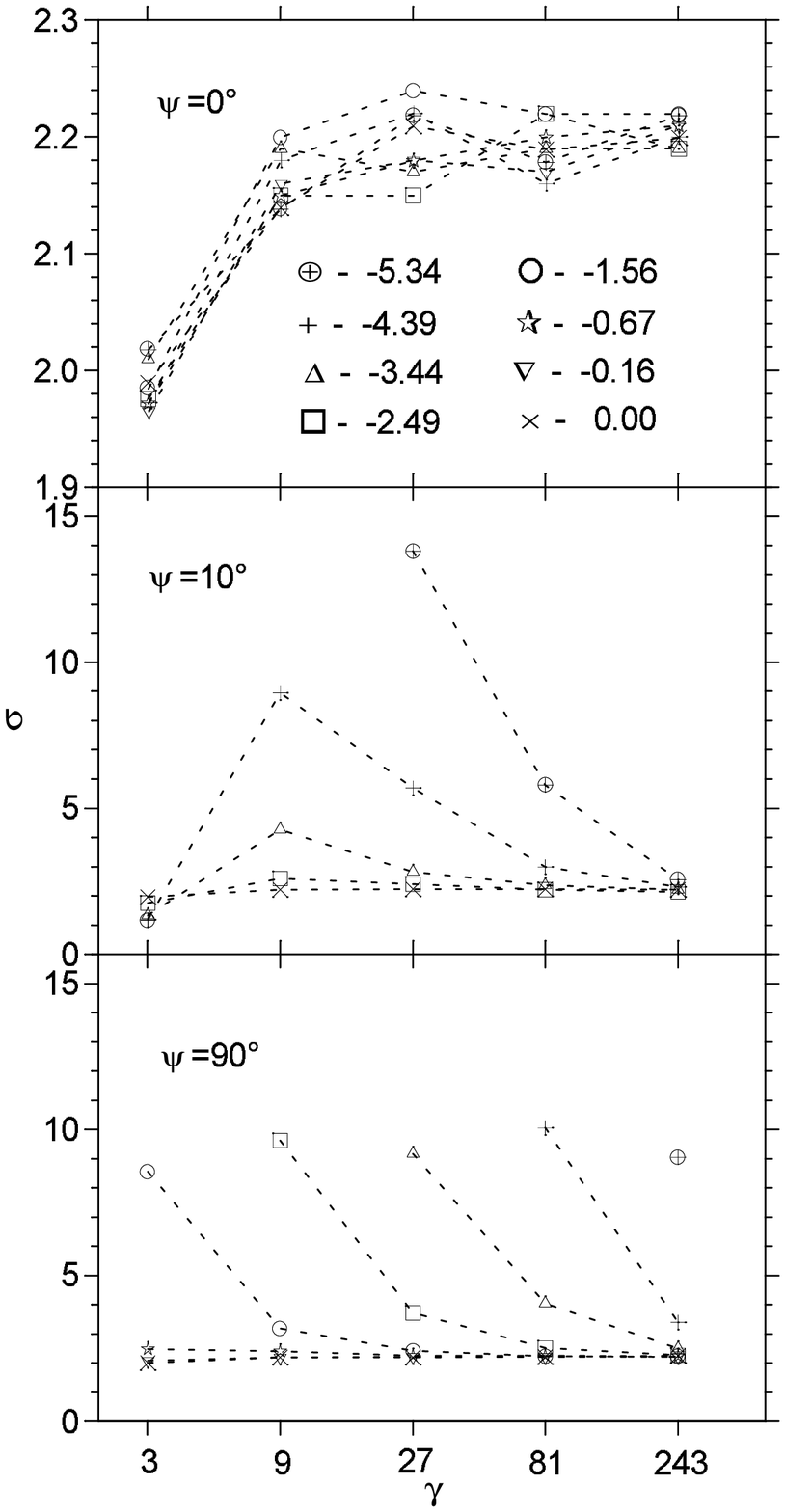}
\caption{   
The simulated spectral indices $\sigma$ for particles accelerated at   
shocks with different Lorentz factors $\gamma$. Results for a given   
$\kappa_\perp / \kappa_\| $ are joined with lines; the respective value   
of $\log_{10} \kappa_\perp / \kappa_\| $ is marked by the point shape
(see upper panel). The results for different magnetic field inclinations
$\psi$ are given in the successive panels: (a) $\psi = 0^\circ$,
(b) $\psi = 10^\circ$, and (c) $\psi = 90^\circ$.}
\label{fig12}
\end{figure}

\begin{figure}
\vspace*{7.0cm}
\includegraphics{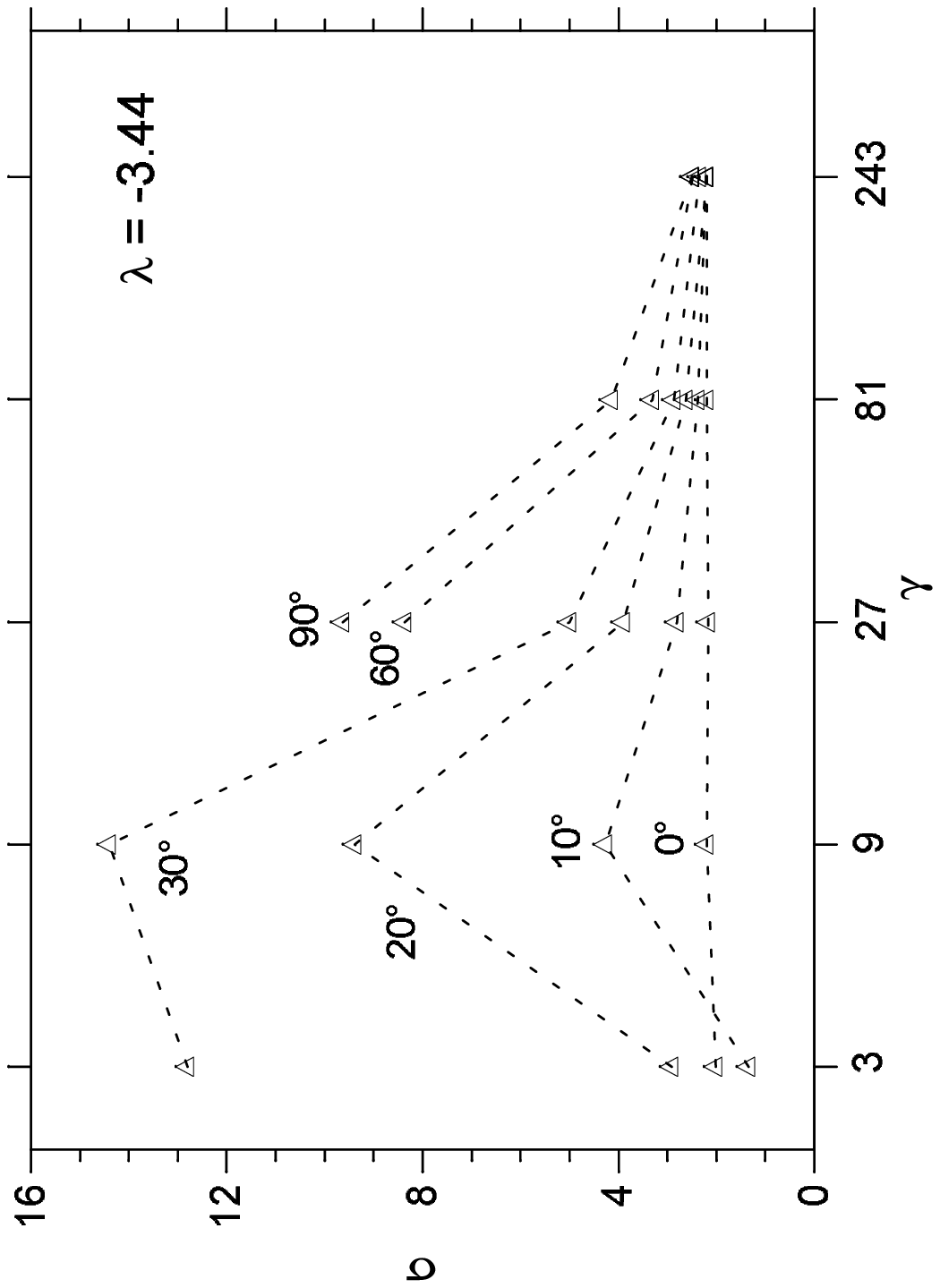}
\caption{
The simulated spectral indices $\sigma$ for particles accelerated at   
shocks with different Lorentz factors $\gamma$. Results for a given   
upstream magnetic field inclination $\psi$ are joined with dashed   
lines; the respective value of $\psi$ is given near each curve. The   
value $\lambda \equiv \log_{10} (\kappa_\perp / \kappa_\| )$ is given
in the   
figure.}
\label{fig13}
\end{figure}

In successive panels in Fig.~4.2 the energy spectral indices, $\sigma$,
for varying $\psi$ and $\kappa_\perp/ \kappa_\|$ are presented. For a
parallel shock ($ \psi = 0^\circ $) the amount of scattering does not
influence the spectral index and for the growing $\gamma$ it approaches
$\sigma_\infty \simeq 2.2$. One may note that essentially the same
limiting value was anticipated for the large-$\gamma$ parallel shocks
by Heavens \& Drury (1988). Let us remember that the results for
$\psi = 10^\circ$ are for superluminal shocks if $\gamma > 5.75$. 
In this case, when we go from the `slow' $\gamma = 3$ shocks to higher
$\gamma$ ones, at first the spectrum inclination increases ($\sigma$
grows) but at large $\gamma$ the spectrum flattens to approach the
asymptotic value close to 2.2. The spectrum steepening phase is more
pronounced for small amplitude perturbations (small $\kappa_\perp /
\kappa_\|$), but even at very low turbulence levels the final range of
the spectrum flattening is observed. For larger $\psi$ the situation
does not change considerably, but the phase of spectrum steepening
is wider involving larger values of $\sigma$ and starting at smaller
velocities below the lower limit of our considerations (there may be
no such range involving the steepening phase if the required
velocity is below the sound velocity).

The spectral indices for different magnetic field inclinations, but
for the same value of $\log_{10} ( \kappa_\perp / \kappa_\| ) = -3.44$
are presented at Fig.~4.3. The large spectral indices occurring in the
steepening phase are usually interpreted as a spectrum cutoff. In this
case the main factor increasing the particle energy density is a
nonadiabatic compression in the shock (Begelman \& Kirk 1990).

The particle angular distributions $F(\mu)$ in the $\gamma\gg 1$
shocks can be extremely anisotropic when considered in the upstream
plasma rest frame. However, when presented in the shock rest frame
the distribution is always `mildly' anisotropic. This feature is
illustrated in Fig.~4.4 when $\gamma$ equals 3 or 27 (note that in
Figs.~4.4~-~4.7 the area below each curve is normalized to 100). In
the simulations we observed an interesting phenomenon accompanying
previously discussed spectrum convergence to the limiting inclination:
spectra close to the limit exhibit similar angular distributions at
the shock {\it as measured in the shock rest frame} (Fig.~4.5).

\begin{figure}
\vspace*{14.5cm}
\includegraphics{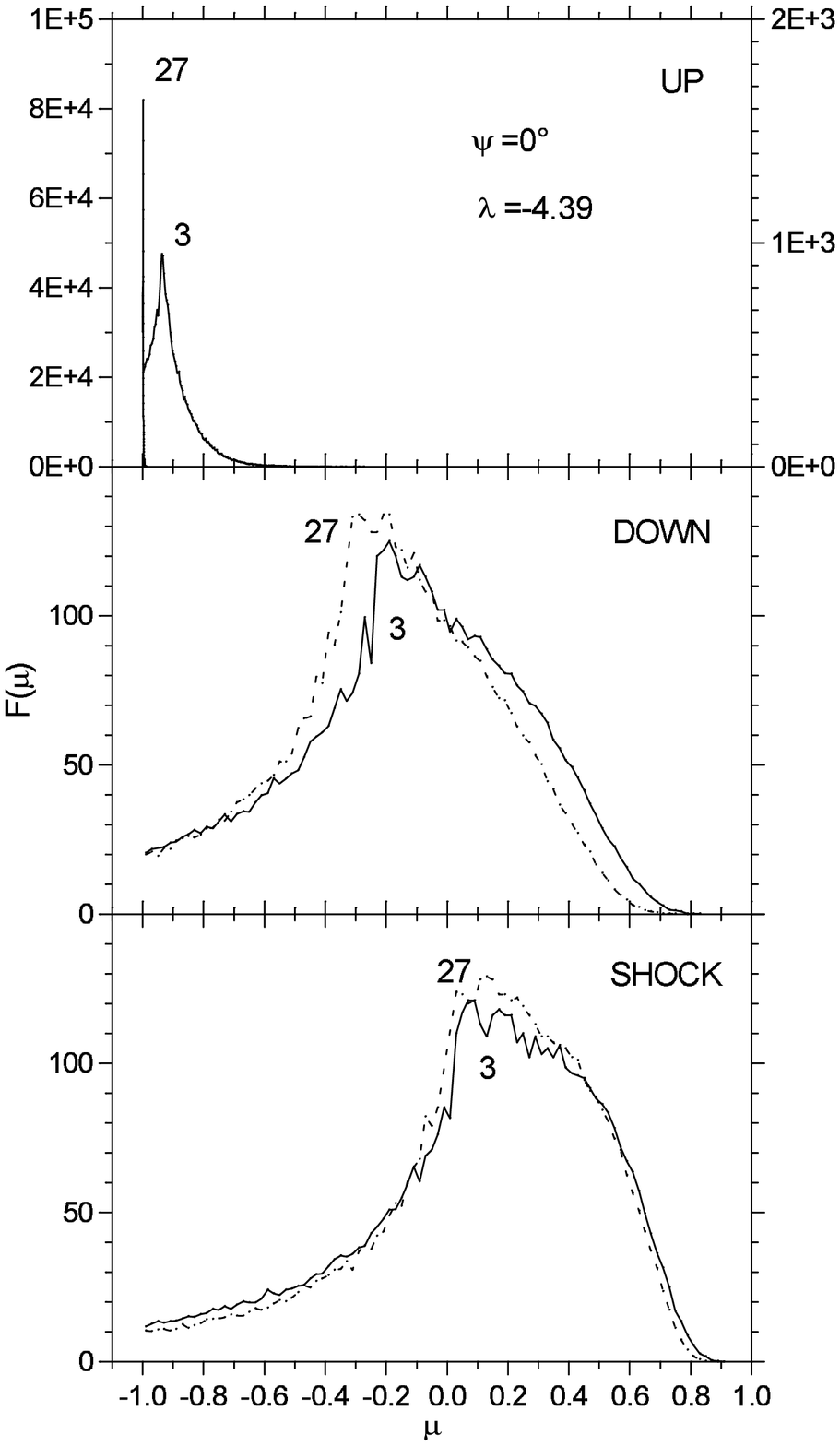}
\caption{
The simulated particle angular distributions in the shock in different
coordinate frames: UP -- the upstream plasma rest frame, DOWN -- the
downstream plasma rest frame and SHOCK -- the shock rest frame. The
results are presented for parallel shocks with the Lorentz factors
27 and 3 given near the respective curves. In the upper panel the left
axis is for $\gamma = 27$ and the right one for $\gamma = 3$.}
\label{fig14}
\end{figure}
\begin{figure}
\vspace*{7.0cm}
\includegraphics{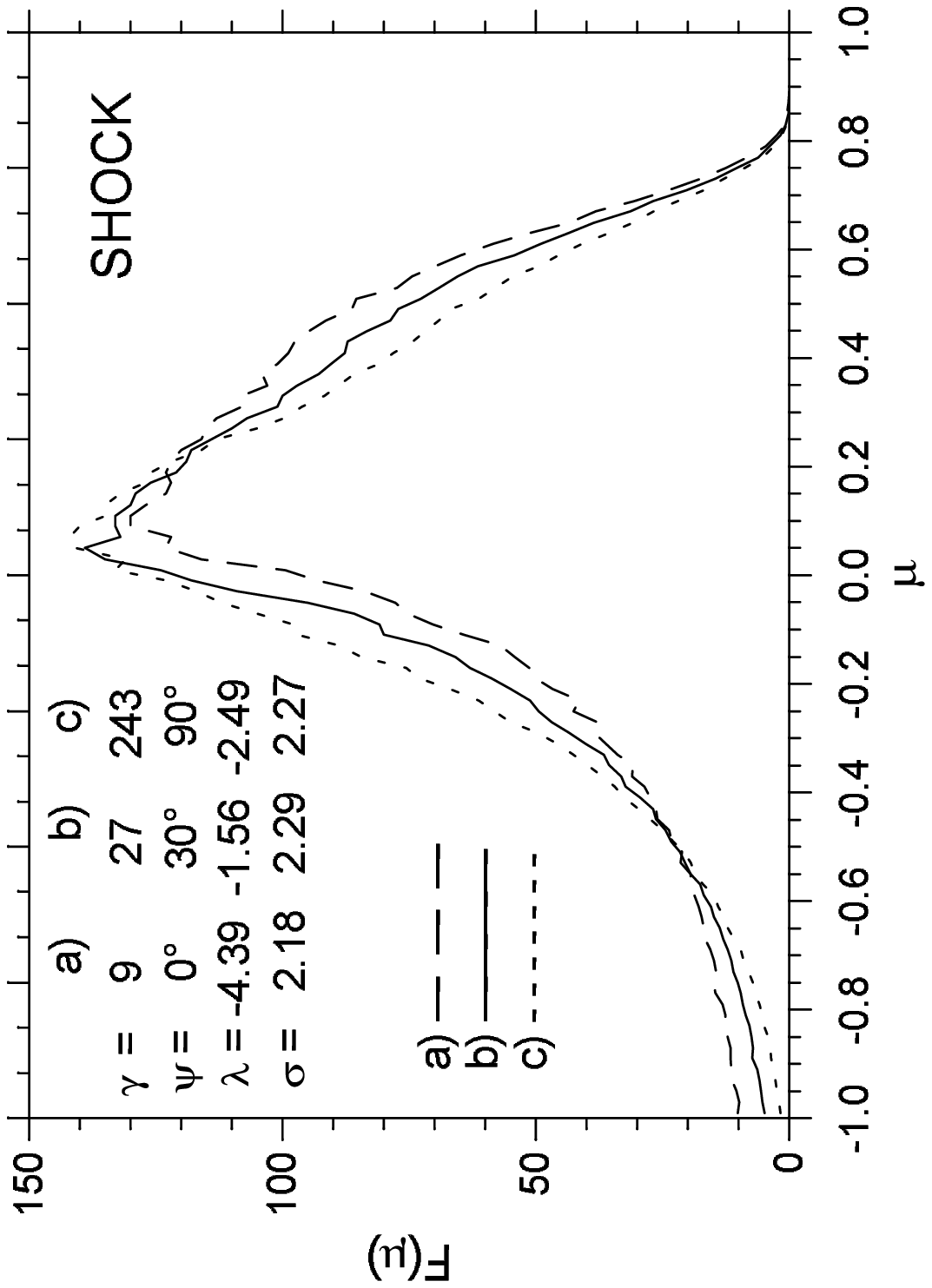}
\caption{
Examples of the shock rest frame  particle angular distributions 
for different cases with $\sigma$ close to $\sigma_\infty$.}
\label{fig15}
\end{figure}
\begin{figure}
\vspace*{7.0cm}
\includegraphics{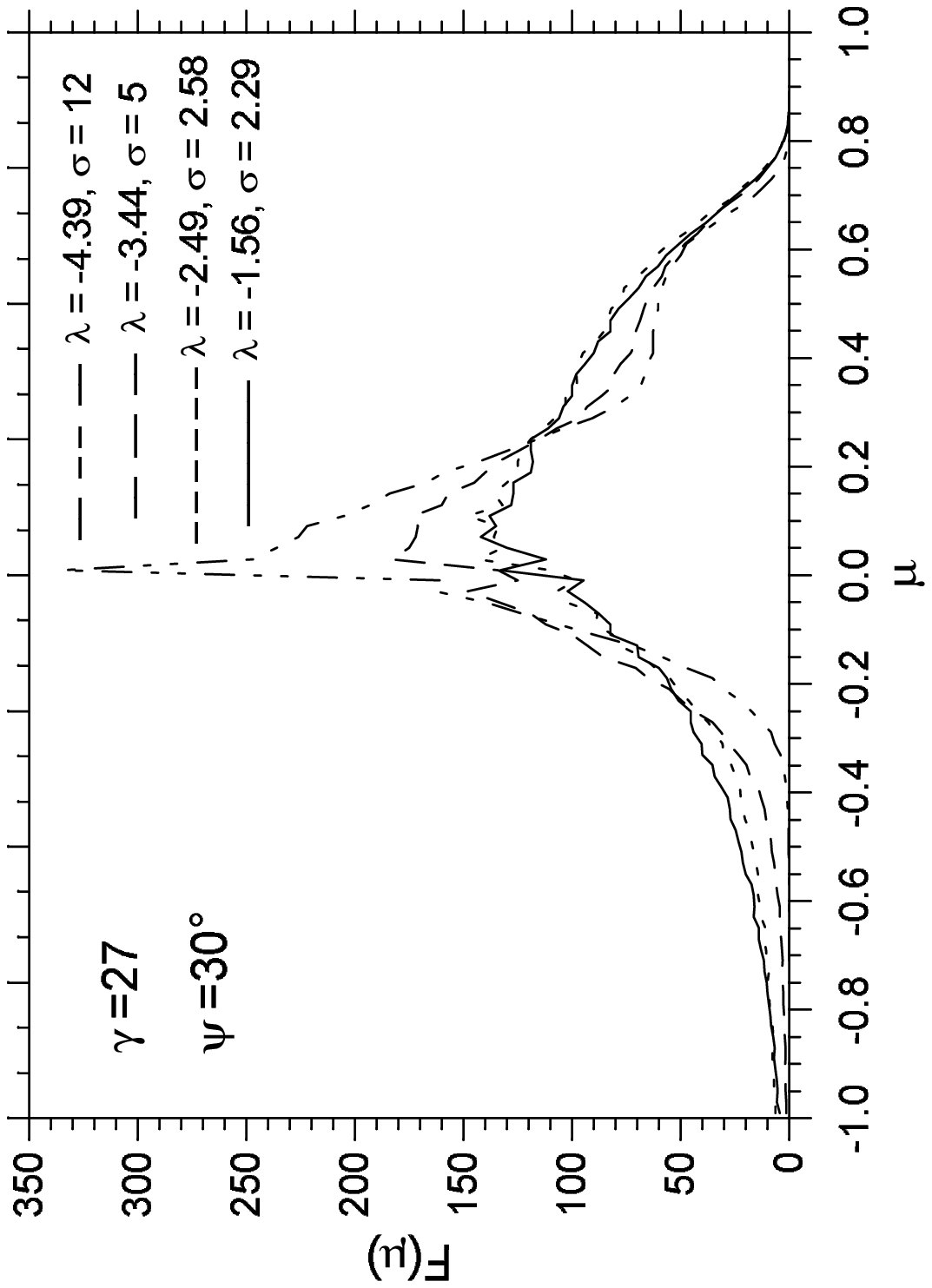}
\caption{
The shock rest frame  particle angular distributions for $\gamma = 
27$ and  $\psi = 30^\circ$. Curves are presented for increasing 
$\lambda \equiv \log_{10} \kappa_\perp / \kappa_\| $ and $\sigma$
approaching $\sigma_\infty$. The last curve is the same as the curve
(b) at Fig. 4.5}
\label{fig16}
\end{figure}
\begin{figure}
\vspace*{7.0cm}
\includegraphics{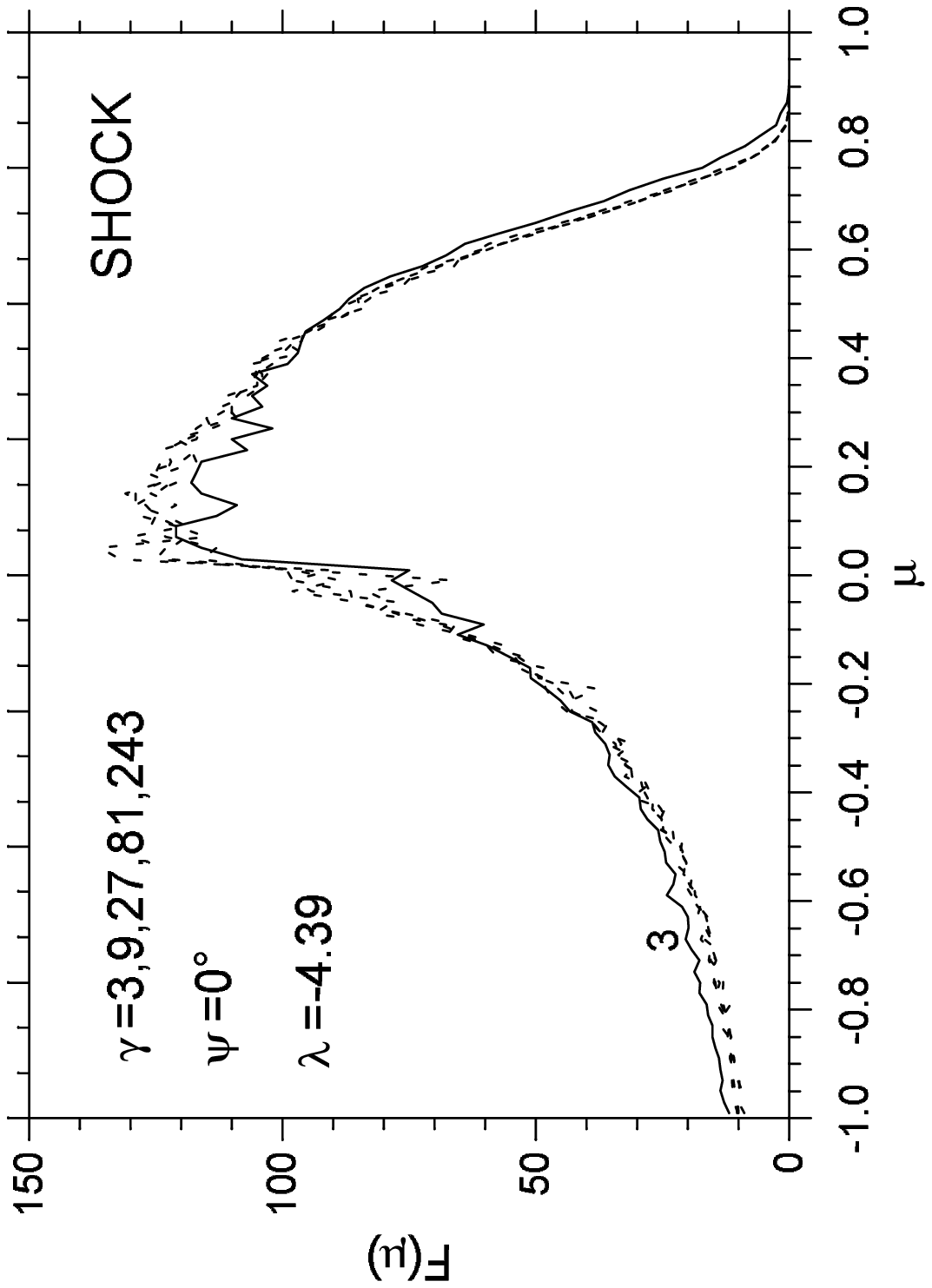}
\caption{
The shock rest frame  particle angular distributions for
parallel shocks with $\gamma = 3$, $9$, $27$, $81$, $243$.
A visible deviation of the distribution for $\gamma = 3$ (full line)
results from the slightly larger compression occurring in such a shock.}
\label{fig17}
\end{figure}
Again, this feature is independent of the background conditions, and
the difference between the actual angular distribution and the limiting
one reflects the difference between the spectral index $\sigma$ and
$\sigma_\infty$~(cf. Fig.~4.6). For parallel shocks with $\gamma \ge 9$
where the spectral index is essentially constant
$\sigma = \sigma_\infty$ this distribution is independent of the value
of $\gamma$ and the perturbation amplitude $\kappa_\perp / \kappa_\|$
(Fig.~4.7).

For large $\gamma$ shocks we observe the convergence of the derived
energy spectral indices to the value $\sigma_\infty \approx 2.2$
independent of the background conditions. This unexpected result
provides a strong constraint for the acceleration process in large
$\gamma$ shocks and it can be quantitatively explained with arguments
presented at the beginning of this chapter. Our interesting finding
do not fully explained with such simple arguments is of the belief
that the resulting spectral index is the same for oblique and  parallel
shocks. Our derivations are limited to the test particle approach.
However, as the obtained spectra are characterized with $\sigma > 2.0$
any nonlinear back reaction effects are not expected to affect the
acceleration process within the spectrum high energy tail with
$\sigma \approx \sigma_\infty$.

\section{The acceleration time scale}

\begin{figure}
\vspace*{7.0cm}
\includegraphics{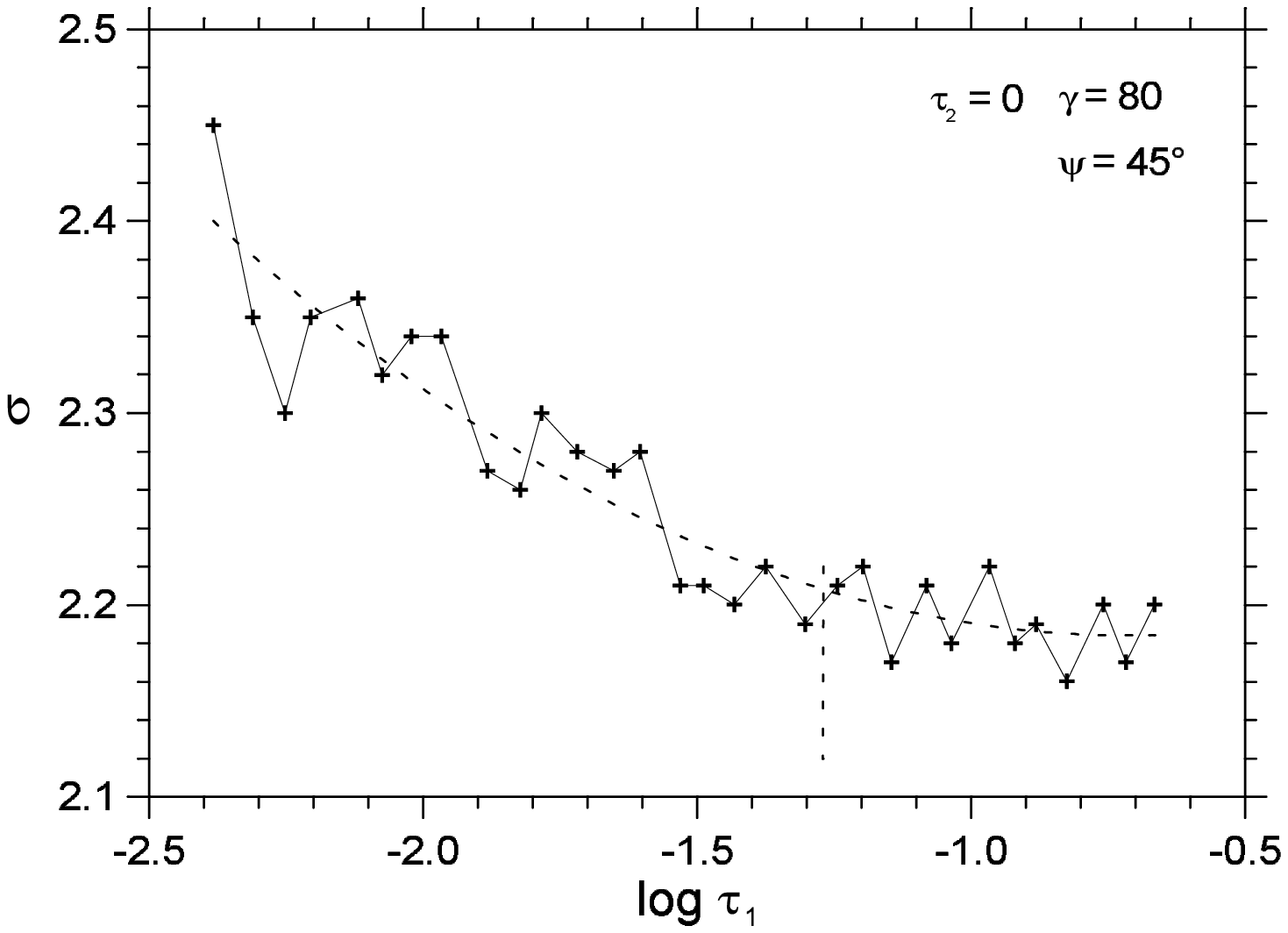}
\caption{
Simulated spectral indices as a function of magnetic field fluctuations
upstream of the shock. Fluctuations downstream of the shock are neglected.
The chosen $\tau_{1}$ value for $\gamma=80$ and $\psi=45^\circ$ is
pointed by a dashed line. A second-degree polynomial fit is also marked by
a dashed line.}
\label{fig18}
\end{figure}
In the following simulations we consider shocks with $\gamma=$
20, 40, 80, 160, 320, magnetic field inclinations $\psi=$ $15^\circ$, 
$30^\circ$, $45^\circ$, $60^\circ$, $75^\circ$, $90^\circ$ and downstream
values of magnetic field fluctuations $\tau_{2}= 0$, $1.0\cdot 10^{-3}$, 
$1.1\cdot 10^{-2}$, $0.11$, $0.69$.

Simulations prove that fluctuations upstream of the shock (measured by
$\tau_{1}$, $\tau\equiv\kappa_\perp / \kappa_\| $) and downstream of the
shock (measured by $\tau_{2}$) influence the acceleration process
independently. The minimum fluctuations upstream of the shock needed to
run the acceleration process efficiently tend to zero when
$\gamma \rightarrow \infty$. We checked by simulations with different
$\tau_{2}$ that its value does not influence the spectral index
considerably for any given $\tau_{1}$ with exception of only the injection
phase of the upstream isotropic distribution.

Thus, as a first case we consider downstream conditions without magnetic
field fluctuations. By simple data inspection (cf. Fig.~4.8) we look for
minimum $\tau_{1}$ where the spectral index reaches its limit of 2.2 and
we apply this value in further simulations. The relation between
$\tau_{1}$, $\gamma$ and $\psi$ can be roughly fitted with the equation
$\tau_{1} = 0.25 \, \gamma^{-1.2} \, \psi$ in the considered range of
shock parameters. We repeated simulations for a number of cases with
different $\gamma$ and $\psi$ and $\tau_{2}\not=0$. The obtained results
are in good agreement with the ones derived from the above equation up
to $\tau_{2}= 0.11$.

\begin{figure}
\vspace*{14.5cm}
\includegraphics{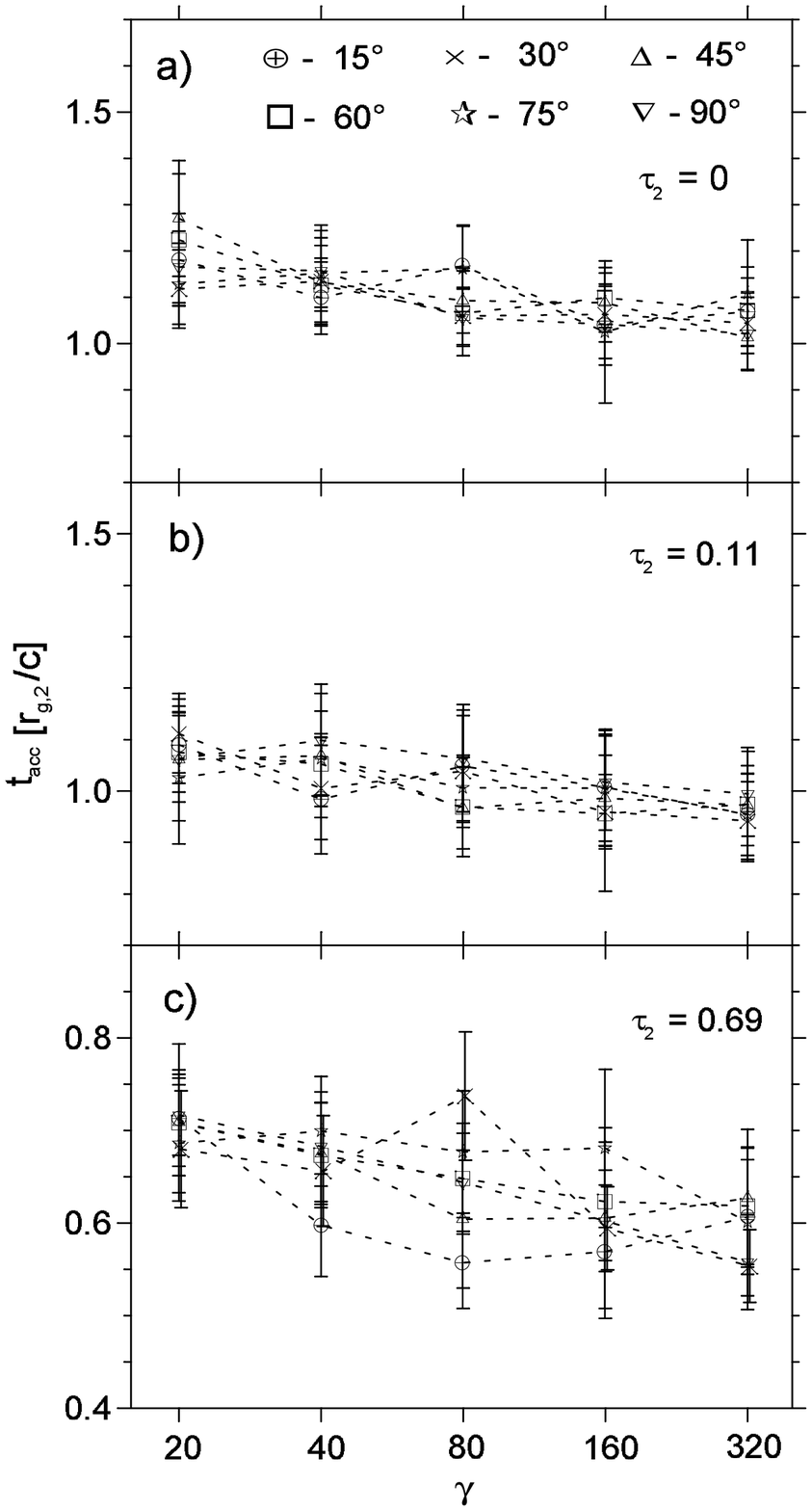}
\caption{
The simulated acceleration time as a function of the shock Lorentz factor:
a) without fluctuations downstream of the shock, b) with fluctuations
downstream of the shock, c) fluctuations downstream of the shock dominate
homogeneous magnetic field. Results for a given upstream magnetic field
inclinations given in panel a) are joined with dashed lines.}
\label{fig19}
\end{figure}

Values of the acceleration time $t_{acc}$ for three amplitudes of magnetic
field fluctuations downstream of the shock are presented in Fig. 4.9.
In the figure one can see the lack of change of  $t_{acc}$ with $\psi$ but
it slowly decreases to the asymptotic value with $\gamma$. In the simulations
we have observed the tendency of $t_{acc}$ to grow when $\sigma$ increases up
to 2.3-2.4 and no further change if magnetic field fluctuations upstream of
the shock grow. For $\tau_{2}\leq0.11$ the asymptotic value of the
acceleration time is close to $r_{g}/c$. It occurs that $r_{g}/c$ is a good
unit provided that the homogeneous magnetic field dominates the randomly
component. Unfortunately, when this condition fails the meaning of $t_{acc}$
becomes unclear in the simulations then. For this reason we will not discuss
further the case of $\tau_{2}$=0.69.

Approximate calculations of Gallant \& Achterberg (1999) showed that
$t^{U}_{U}/t^{D}_{U}\simeq 1$, where $t^{U}_{U}$ is the particle mean residence
time upstream of the shock (upper index) as measured in the upstream plasma
rest frame (lower index), and D in $t^{D}_{U}$ stands for the downstream
residence time. However, they were not able to consider the anisotropic
particle momentum distribution and our results in Fig. 4.10 {\it transformed to
the upstream plasma rest frame} with $t^{U}_{U}/t^{D}_{U}$ within the range
$0.01-0.1$ are more adequate for real situations. Additionally, the above
authors applied an extremely irregular magnetic field upstream of the shock
represented by randomly oriented magnetic cells with field amplitude $B$ and
they measured time in the upstream unit of $r_{g}(B)/c$. As a result they
obtained that $t^{U}_{U}/t^{D}_{U}$ could be much larger than 1 in the case.

\begin{figure}
\vspace*{7.35cm}
\includegraphics{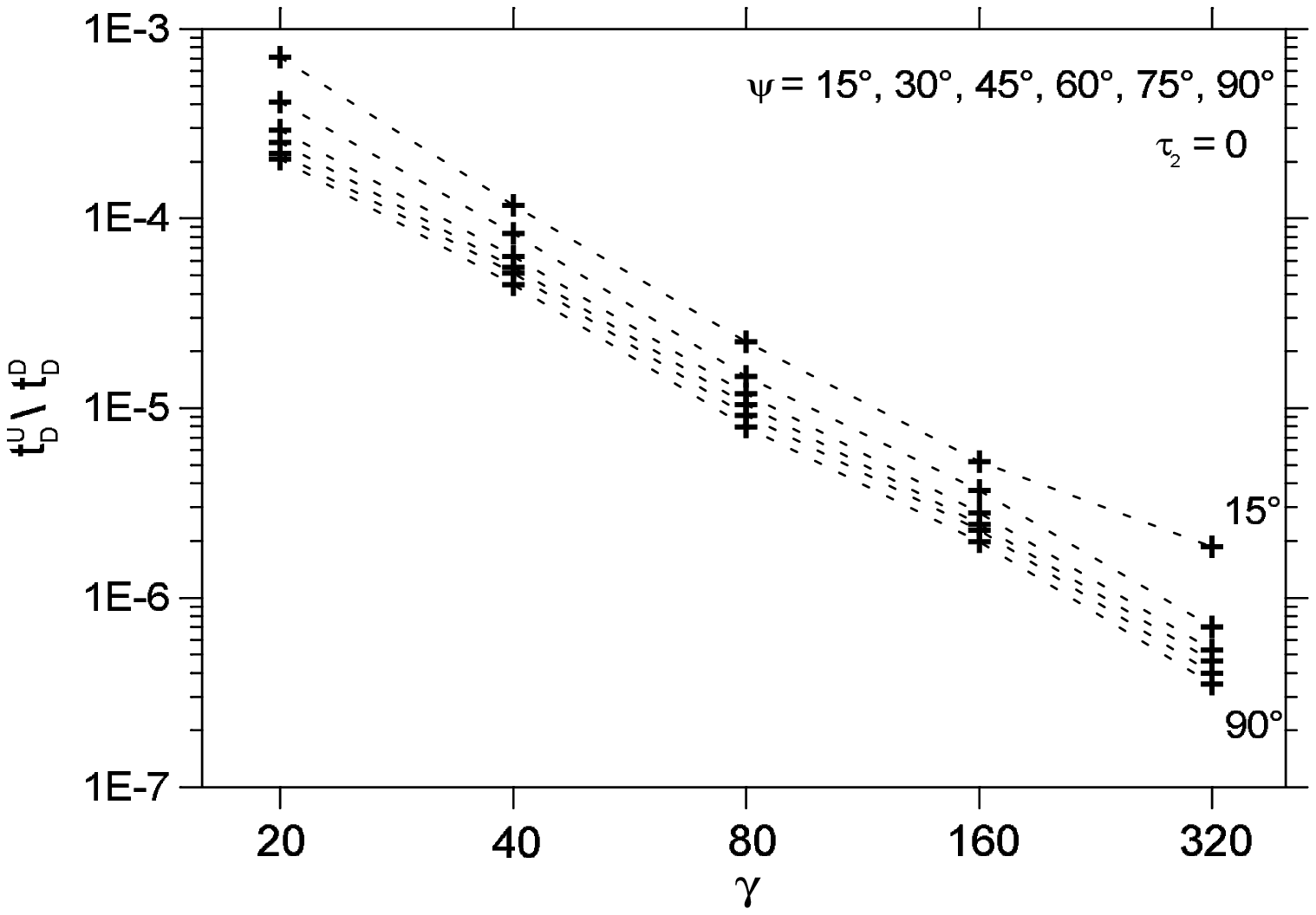}
\vspace*{0.1cm}
\caption{
The ratio of the mean time a particle spends upstream of the shock to the time
it spends downstream of the shock as a function of the shock Lorentz factor.
It slightly decreases with growing upstream magnetic field inclination. Dashed
lines join points with a constant $\psi$. Apparent deviation of the point with
$\gamma=320$, $\psi=15^\circ$ is real.}
\label{fig20}
\end{figure}

Just before the spectral index reaches its minimal value (cf. Fig.~4.8)
$\Delta \Omega_{S}$ stabilizes near the limit which value does not further depend
on the magnetic field inclination as is seen in Fig. 4.11. Momentum vectors of
particles crossing downstream of the shock have similar distributions as measured
in the downstream plasma rest frame if $\Delta \Omega_{S}$ approaches the maximum
value. Then, it follows that parameters we consider below depend only on
$\tau_{2}$.

\begin{figure}
\vspace*{7.0cm}
\includegraphics{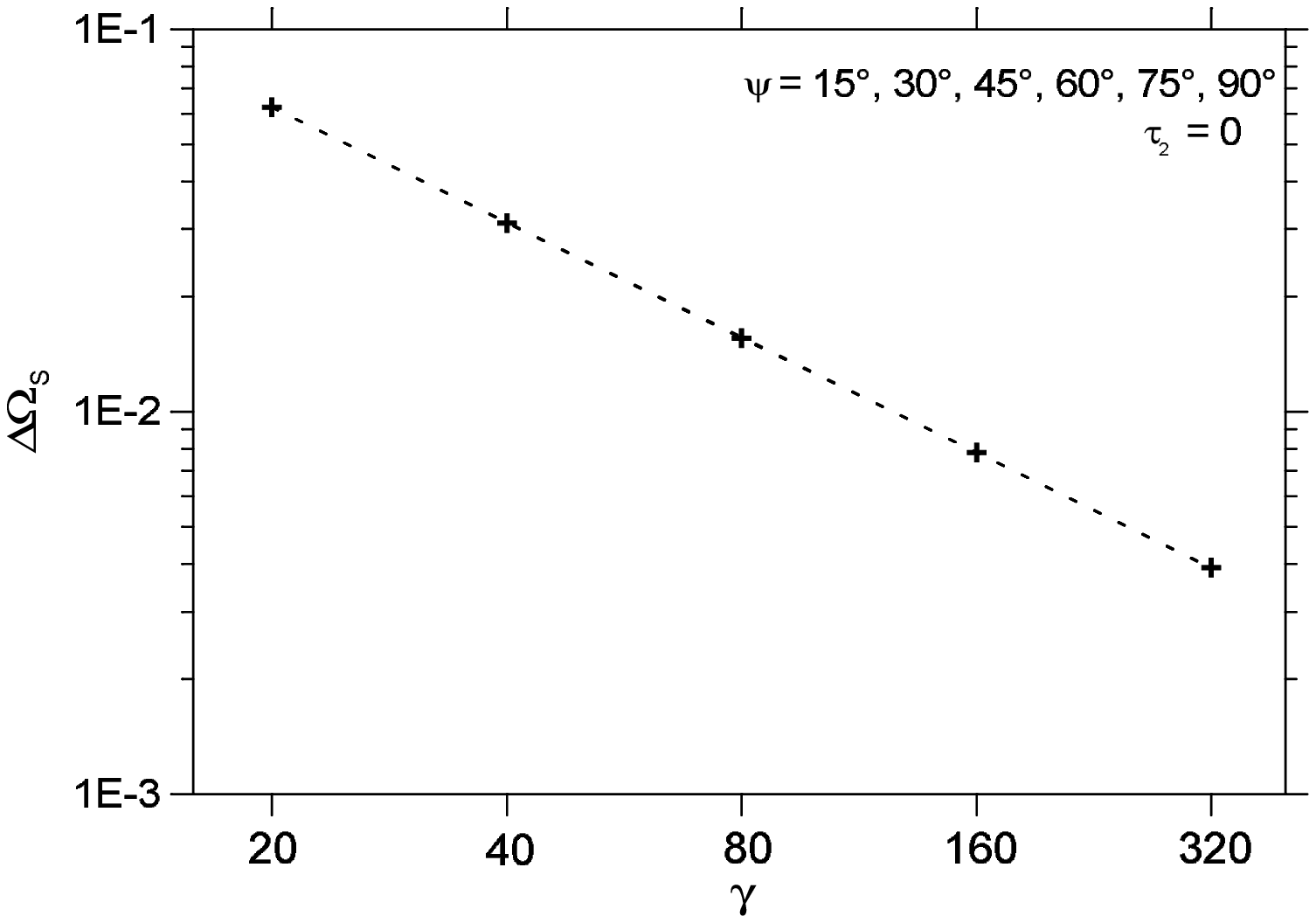}
\caption{
The mean deflection angle resulting from scattering in an inhomogeneous magnetic
field upstream of the shock as a function of the shock Lorentz factor
$\gamma$ with $\tau_{1}$ chosen in a way as illustrated in Fig. 4.8. Six dashed
lines for different $\psi$ are identical.}
\label{fig21}
\end{figure}

\begin{figure}
\vspace*{9.2cm}
\includegraphics{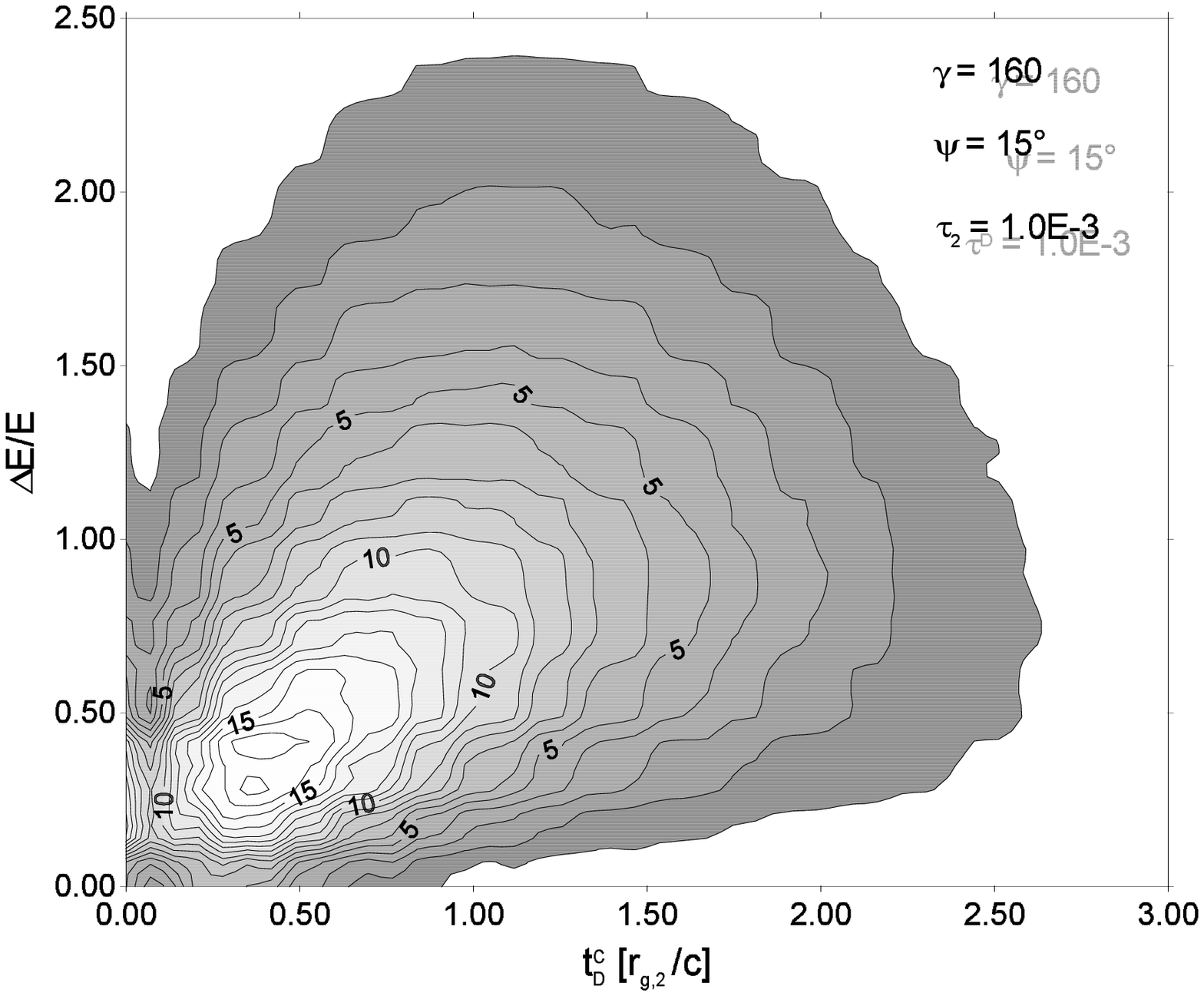}
\caption{
Distribution of particle energy gains $\Delta E/E$ as measured in the downstream
plasma rest frame during one cycle `C' downstream-upstream-downstream versus the
time $t^{C}_{D}$[$r_{g,2}/c$]. The shock parameters are given in the picture.
The mean energy gain $\langle\Delta E/E\rangle_{D}$ grows with $t^{C}_{D}$ up to
$t^{C}_{D}\simeq 2$ and decreases afterwards.}
\label{fig2}
\end{figure}

For growing $\tau_{2}$ ($\tau_{2}=0, 1.0\cdot 10^{-3}, 1.1\cdot 10^{-2}, 0.11$)
\footnote{Below, we provide the respective series of a given simulated parameter
for this sequence of $\tau_{2}$.} the acceleration time is constant and
accompanied by a slow increase of the mean energy gain in one cycle
downstream-upstream-downstream
$\langle\Delta E/E\rangle_{D} = 0.89,\, 0.94,\, 1.0,\, 1.1$, and a slight
decrease of the fraction of particles that reach the shock again after crossing
it downstream, $\langle\Delta n/n\rangle = 0.51,\, 0.50,\, 0.48,\, 0.44$.
Simultaneously the mean time a particle spends downstream of the shock grows as
$t^{D}_{D} = 0.96,\, 1.0,\, 1.2,\, 1.35$. Time that a particle spends upstream
of the shock can be neglected in this rest frame as is visible in Fig. 4.10. It
implies, approximately,
$$
t_{acc}=t_{D}^{D}/\langle\Delta E/E\rangle_{D} \eqno(4.1)
$$
if one neglects correlations between these quantities (cf. Fig.~4.12).
Similarly we can roughly estimate the value of the energy spectral
index of accelerated particles as
$$
\sigma\simeq 1-\ln(\langle\Delta n/n\rangle)/
\ln(\langle\Delta E/E\rangle_{D}+1) \qquad . \eqno(4.2)
$$

\section{The acceleration through particle reflection}

Particles with an initial momentum $p_0$ taken as the momentum unit,
$p_{0}=1$, were injected  at the distance of $2r_{g}$ ($r_{g}$ -
particle gyroradius) upstream of the shock front. For all particles
we derived their trajectories until crossing the shock downstream, and
then upstream, or were advected with the downstream plasma, to reach a
distance of $4 r_g$ downstream of the shock. For each single particle
interaction with the shock the particle momentum vector was recorded
so we were able to consider angular and energy distributions of such
particles. We considered shocks with Lorentz factors $\gamma= 10$,
$160$ and $320$. For each shock we discussed the acceleration
processes in conditions with the magnetic field inclinations
$\psi= 0^\circ$, $10^\circ$, $70^\circ$ and with 16 values for the
turbulence amplitude measured by the ratio $\tau$ of the cross-field
diffusion coefficient $\kappa_\perp$ to the parallel diffusion
coefficient $\kappa_\|$. The applied values of $\tau$ were taken from
the range of ($3.2\cdot 10^{-6}$, $0.95$), approximately uniformly
distributed in $\log \tau$~. In each simulation run we derived
trajectories of $5\cdot 10^{4}$ particles with the initial momenta
isotropically distributed in the upstream rest frame.

In the downstream plasma rest frame the shock moves with velocity
$c/3$. This velocity is comparable to the particle velocity $c$.
Therefore, from all particles crossing the shock downstream only the
ones with particular momentum orientations will interact with the shock
again; the remaining particles will be caught in
the downstream plasma flow and advected far from the shock front. In the
simulations we considered this process quantitatively. However, let us
first present a simple illustration.

Large compression ratios occurring in ultrarelativistic shocks, as
measured between the upstream and downstream plasma rest frames, lead
for nearly all oblique upstream magnetic field configurations to the
quasi-perpendicular configurations downstream of the shock. Thus, let us
consider for this illustrative example a shock with a non-perturbed
perpendicular downstream magnetic field distribution. Particle crossing
the shock downstream with inclination to the magnetic field $\vartheta$
and the phase $\varphi$ -- both measured in the downstream plasma rest
frame, $\varphi = \pi/2$ for a particle velocity normal to the shock and
directed downstream -- will be able to cross the shock upstream only if
the equation
$$
{c \over 3} \, t = r_{\rm g} \left[ \cos (\varphi + \omega_{\rm g} t ) 
- \cos \varphi \right] \eqno(4.3)
$$
\begin{figure}
\vspace*{7.0cm}
\includegraphics{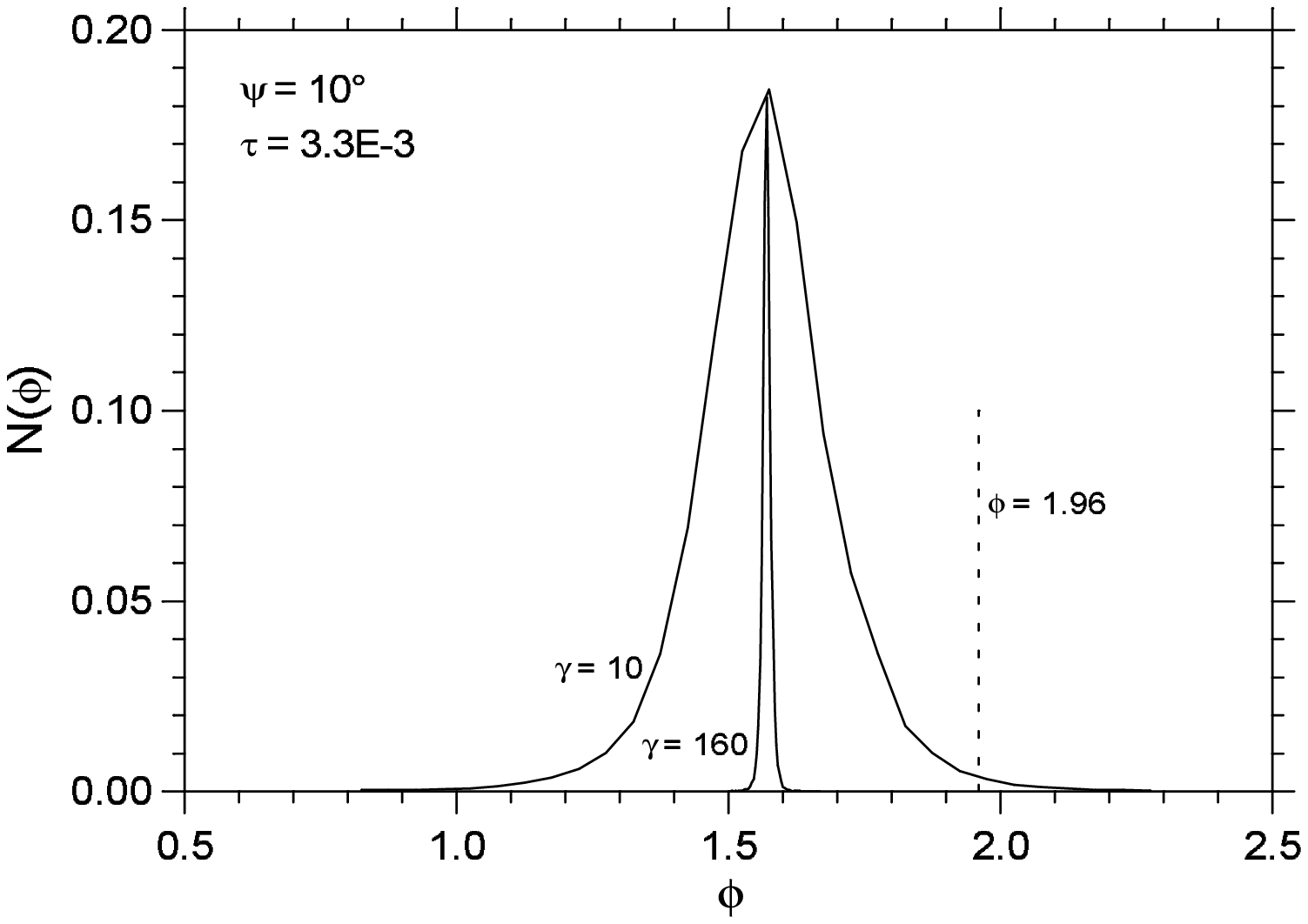}
\caption{A distribution of particle phases for particles crossing 
the shock downstream (as measured in the downstream plasma rest frame), 
if their upstream distribution was isotropic. 
A dashed line delimits a range of particle phases below which particles
are not able to reach the shock again at the perpendicular uniform 
downstream magnetic field. }
\label{fig22}
\end{figure}
\begin{figure}
\vspace*{7.0cm}
\includegraphics{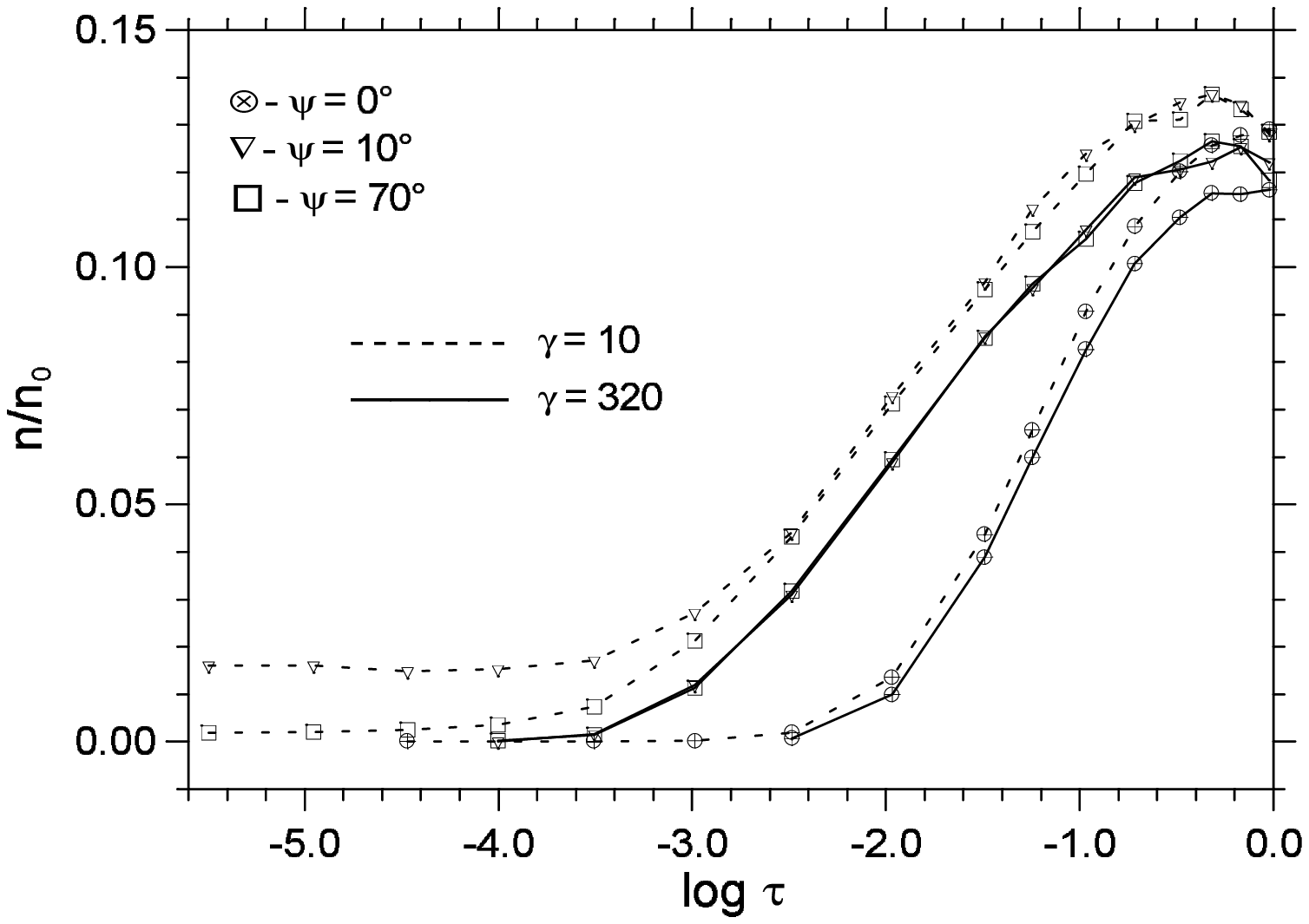}
\caption{
A ratio of the number of reflected particles, $n$, to all particles 
crossing the shock downstream, $n_0$, as a function
of the magnetic field fluctuations amplitude,~$\tau$.}
\label{fig23}
\end{figure}
\clearpage
\begin{figure}[t]
\vspace*{7.0cm}
\includegraphics{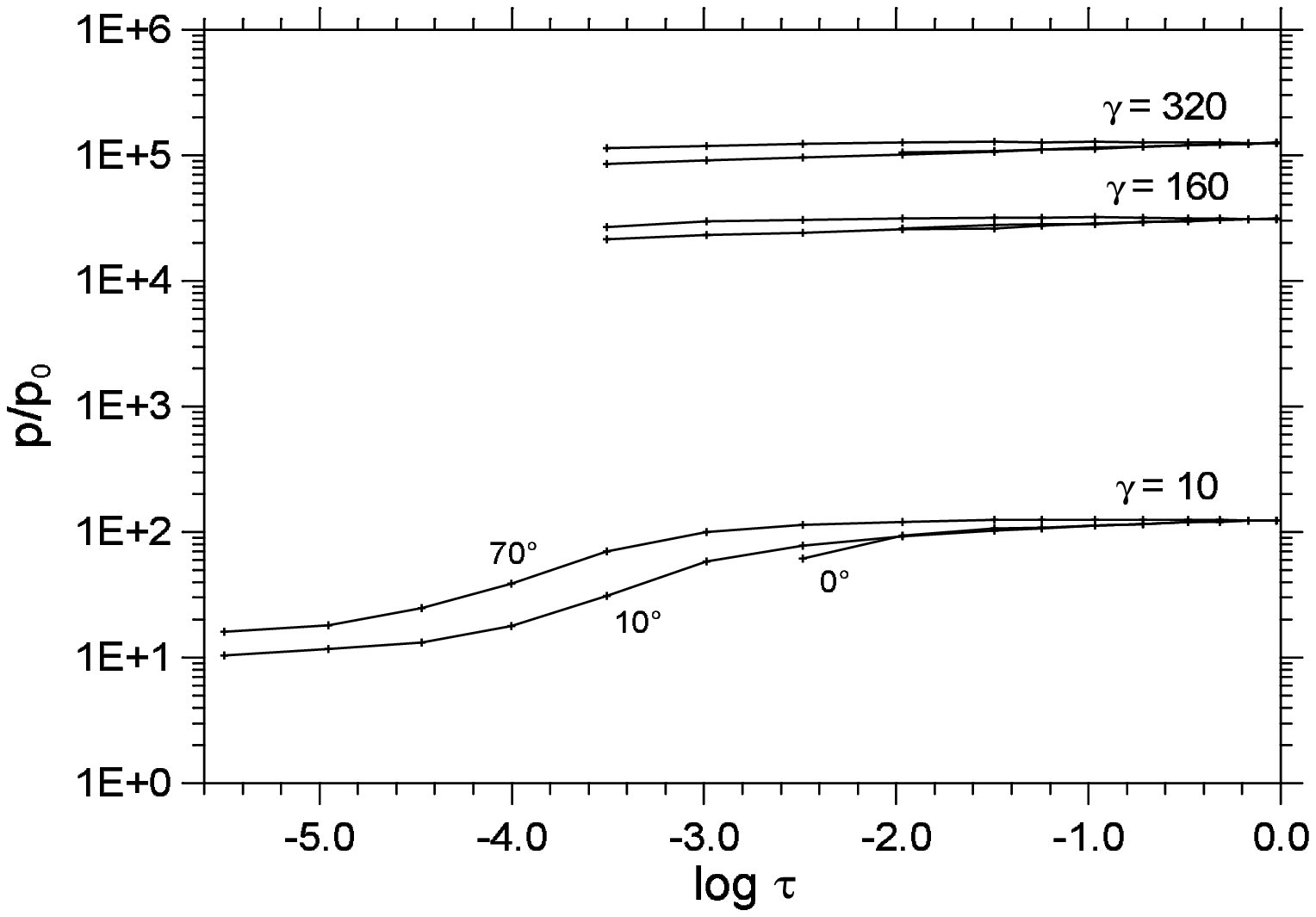}
\caption{
Momentum gains of reflected particles, $p/p_0$, as a function of the
magnetic field fluctuations amplitude $\tau$. For large magnetic field
fluctuations the momentum gain approaches $\approx 1.2 \, \gamma^{2}$
independently of the shock Lorentz factor.}
\label{fig24}
\end{figure}
\noindent
has a solution at positive time $t$. Here $r_{\rm g} = { pc \over eB}
\sin \vartheta$ is the particle gyroradius, $\omega_{\rm g} = {e B \over
p}$ is the gyration frequency, and other symbols have the usual meaning.

An angular range in the space ($\vartheta$, $\varphi$) enabling particles
crossing the shock downstream to reach the shock again can be
characterized for illustration by three values of $\vartheta$. Particles
with $\sin \vartheta = 1$ are able to reach the shock again if $\varphi
\in$(1.96, 3.48), with $\sin \vartheta = 0.5$ if $\varphi \in$(2.96, 3.87)
and with $\sin \vartheta = 1/3$ only for $\varphi = 4.71$. That means that
all particles with $\varphi$ smaller than 1.96 (Fig.~4.13) are not able to
reach the shock again if fluctuations of the magnetic field downstream of
the shock are not present.

For perturbed magnetic fields some downstream trajectories starting in
the ($\vartheta$, $\varphi$) plane outside the reflection range can be
scattered toward the shock to cross it upstream. We prove it by
simulations presented in Fig.~4.14. One may observe that increasing the
perturbation amplitude leads to an increased number of reflected particles
reaching $\approx 13$\% in the limit of $\tau = 1$. For large magnetic
field fluctuations the mean relative energy gains of reflected particles
are close to $1.2 \gamma^2$ for the shock Lorentz factors considered.
One may note a variation of the energy gain with growing $\tau$.
The points resulting from simulations for the smallest values of $\tau$
were not included in Fig.~4.15 because of the small number of reflected
particles (cf. Fig.~4.14).

\chapter{Summary}
We performed Monte Carlo simulations for
shock waves with parallel and oblique (either, sub-luminal and
super-luminal) magnetic field configurations with different amounts of
scattering along particle trajectories. Field perturbations with
amplitudes ranging from very small ones up to $\delta B \sim B$ are
considered.

In chapter 2 (cf. Bednarz \& Ostrowski 1996) we
demonstrate the existence of correlation between particle energy gains
and its diffusive times. The analogous correlation is expected for the
probability of particle escape downstream the shock. Therefore, for
defining the acceleration time scale we use the rate of change of the
spectrum cut-off momentum which accommodate all such correlations. 

Acceleration times scales in relativistic shocks are discussed in
chapter 3 (cf. Bednarz \& Ostrowski 1996). In parallel shocks
$T_{acc}^{(c)}$ diminishes with the growing perturbation amplitude and
the shock velocity. However, it is approximately constant for the
increasing turbulence level if we use the respective diffusive time
scale as the time unit. Another feature discovered in oblique shocks
is that due to the cross-field diffusion $T_{acc}^{(c)}$  can change
with $\delta B$ in a non-monotonic way. The acceleration process
leading to the power-law particle spectrum in a super-luminal shock
is possible only in the presence of large amplitude turbulence. Then,
the shorter acceleration times occur when the perturbations'
amplitudes are smaller and the respective spectra steeper. We
discussed the coupling between the acceleration time scale and the
particle spectral index in oblique shock waves with various field
inclinations and revealed a possibility for non-monotonic relations of
these quantities. The shortest acceleration time scales seen in the
simulations are below the particle gyroperiod upstream of the shock.
These times do not require the ultrarelativistic shock velocities,
but may occur in mildly relativistic ones with the quasi-perpendicular
magnetic field configuration. One should note that due to the larger
magnetic field downstream of the shock in this short time the particle
trajectory can follow a few revolutions near the shock with only a
short section of each one penetrating the upstream region.

The presented estimates of the acceleration time scale provide an
interesting possibility for modeling shock waves in the conditions
where the electron spectrum cut-off energy is determined by the
balance of gains and losses. If one is able to derive the respective
acceleration rate from the knowledge of the energy loss process and the
particle spectral index is also known then both these values provide
constraints for the acceleration process which could be further used to
reduce the parameter space available for the considered shock wave
(cf. Fig.~3.4). 

In chapter 4 (cf. Bednarz \& Ostrowski 1998, 1999; Bednarz 1999) we
discussed the acceleration mechanism that holds in ultrarelativistic
shocks. We discovered convergence of spectral indices to the universal
asymptotic value $\sigma_{\infty}\simeq 2.2$ and we considered high
particle anisotropy that accompanies particle acceleration. The
simulations yielded that acceleration time derived from formula
applied in ultrarelativistic shocks approximately equals the time
derived from the formula neglecting correlations (4.1) and the
constant value of 1.0 can be used for $\tau \leq 0.11$.

The presented results are to be applied in models of GRB sources
involving ultrarelativistic shock waves (cf. Bednarz 1999). One should
note that the mean downstream plasma proton energies can reach there
several tens of GeV (cf. Paczy\'nski \& Xu 1994) and the lower limit of
the considered cosmic ray energies has to be larger than this scale.
For shocks propagating in (e$^-$, e$^+$) plasma the involved thermal
energies are lower, $\sim \gamma$ MeV. These estimates provide the
respective lower limits for the accelerated cosmic ray particles. For
the physical conditions considered in GRB sources the acceleration
process can provide particles with much larger energies limited only by
the condition that the energy loss processes (radiative, or due to
escape) are ineffective in the downstream gyroperiod time scale. We note
a striking coincidence of our limiting spectral index with the value
derived for energetic electrons from gamma-burst afterglow observations.
Waxman (1997) used a fireball model of GRBs and showed from the
functional dependence of the flux on time and frequency that
$\sigma=2.3\pm0.1$ in the afterglow of GRB 970228. Galama et al. (1998)
made two independent measurements of the electron spectrum index in the
afterglow of GRB 970508 which was very close to $2.2$.

In the end we have shown that efficiency of `$\gamma^2$' reflections in
ultrarelativistic shock waves strongly depends on fluctuations of
magnetic field downstream of the shock. In the most favorable conditions
with high amplitude turbulence downstream the shock the reflection
efficiency is a factor of 10 or more smaller than the values assumed by
other authors. Moreover, due to the magnetic field compression at the
shock we do not expect the required large values of
$\kappa_\perp/\kappa_\parallel$ to occur behind the shock (cf. a
different approach of Medvedev \& Loeb 1999). Therefore, with the actual
efficiency of 1 - 10 \% there is an additional difficulty for models
postulating UHE particle acceleration at GRB shocks (cf. Gallant \&
Achterberg 1999). Let us note, however, that the mean downstream
trajectory of the reflected particle involves only a fraction of its
gyroperiod. Thus the presence of compressive long waves in this region
leading to non-random trajectory perturbations could modify our
estimates.

\chapter{Bibliography}

Appl, S., Camenzind, M., 1988, Astron. Astrophys. {\bf 206}, 258\\
Axford W. I., Leer E., Skadron G., 1977,
  in Proc. 15th Int. Cosmic Ray Conf. {\bf 11},
  \hspace*{0.4truecm} p. 132, Plovdiv\\
Ballard K. R., Heavens A. F., 1991, Mon. Not. R. Astron. Soc. {\bf 251}, 438\\
Ballard K. R., Heavens A. F., 1992, Mon. Not. R. Astron. Soc. {\bf 259}, 89\\
Bednarz J., 1998, in Proc. 16th European Cosmic Ray Symposium, p. 287,
  Alcal\'a
  \hspace*{0.4truecm} de Henares\\
Bednarz J., 1999, submitted\\
Bednarz J., Ostrowski M., 1996, Mon. Not. R. Astron. Soc. {\bf 283}, 447\\
Bednarz J., Ostrowski M., 1997a, in Proc. Conf. Relativistic Jets in AGN's,
  p. 158,
  \hspace*{0.4truecm} Cracow\\
Bednarz J., Ostrowski M., 1997b, in Proc. 25th Int. Cosmic Ray Conf. {\bf 4},
  p. 385,
  \hspace*{0.4truecm} Durban\\
Bednarz J., Ostrowski M., 1998, Phys. Rev. Lett. {\bf 80}, 3911\\
Bednarz J., Ostrowski M., 1999, Mon. Not. R. Astron. Soc. {\bf 310}, L11\\
Begelman M. C., Kirk J. G., 1990, Astrophys. J. {\bf 353}, 66\\
Bell A. R., 1978a, Mon. Not. R. Astron. Soc. {\bf 182}, 147\\
Bell A. R., 1978b, Mon. Not. R. Astron. Soc. {\bf 182}, 443\\
Blandford R. D., K\"onigl A., 1979, Astrophys. J. {\bf 232}, 34\\
Blandford R. D., Ostriker J. P., 1978, Astrophys. J. {\bf 221}, L29\\
Decker R. B., 1988, Space Sci. Rev. {\bf 48}, 195\\
de Jager O. C., Harding A. K., Michelson P. F., Nel H. J.,
  Nolan P. L., Sreekumar
  \hspace*{0.4truecm} P., Thompson D. J., 1996, Astrophys. J. {\bf 457}, 253\\
de Hoffman F., Teller E., 1950, Phys. Rev. {\bf 80}, 692\\
Dermer C. D., 1992, Phys. Rev. Lett. {\bf 68}, 1799\\
Chandrasekhar S., 1943, Rev. Mod. Phys. {\bf 15}, 1\\
Chiueh T., Li Z., Begelman M. C., 1998, Astrophys. J. {\bf 505}, 835\\
Ellison D. C., Jones F. C., Reynolds S. P., 1990, Astrophys. J. {\bf 360}, 702\\
Fermi E., 1949, Phys. Rev. {\bf 75}, 1169\\
Galama T. J., Wijers R. A. M. J., Bremer M., Groot P. J., Strom R. G.,
  de Bruyn
  \hspace*{0.4truecm} A. G., Kouveliotou C., Robinson C. R.,
  van Paradijs J., 1998, Astrophys. J. {\bf 500},
  \hspace*{0.4truecm} L101\\
Gallant Y. A., Achterberg A., 1999, Mon. Not. R. Astron. Soc. {\bf 305}, L6\\
Gallant Y. A., Achterberg A., Kirk J. G., 1998, in Proc. 16th European
  Cosmic
  \hspace*{0.4truecm} Ray Symposium, p. 371, Alcal\'a de Henares\\
Gallant Y. A., Arons J., 1994, Astrophys. J. {\bf 435}, 230\\
Heavens A., Drury L'O. C., 1988, Mon. Not. R. Astron. Soc. {\bf 235}, 997\\
Hester J. J. et al., 1995, Astrophys. J. {\bf 448}, 240\\
Hjellming R. M., Rupen M. P., 1995, Nature {\bf 375}, 464\\
Hoyle F., 1960, Mon. Not. R. Astron. Soc. {\bf 120}, 338\\
Hudson P. D., 1965, Mon. Not. R. Astron. Soc. {\bf 131}, 23\\
Hudson P. D., 1967, Mon. Not. R. Astron. Soc. {\bf 137}, 205\\
Jiang D. R., Cao X., Hong X., 1998, Astrophys. J. {\bf 494}, 139\\
Jokipii J. R., 1971, Rev. Geophys. Space Phys. {\bf 9}, 27\\
Jones F. C., Ellison D. C., 1991, Space Sci. Rev. {\bf 58}, 259\\
Kennel C. F., Coroniti F. V., 1984, Astrophys. J. {\bf 283}, 694\\
Kirk J. G., 1988, {\it Habilitation Theses}, preprint No.  345,
  Max-Planck-Institut f\"ur
  \hspace*{0.4truecm} Astrophysik, Garching\\
Kirk J. G., Heavens A., 1989, Mon. Not. R. Astron. Soc. {\bf 239}, 995\\
Kirk J. G., Schneider P., 1988, Astron. Astrophys. {\bf 201}, 177\\
Kirk J. G., Schneider P., 1987a, Astrophys. J. {\bf 315}, 425\\
Kirk J. G., Schneider P., 1987b, Astrophys. J. {\bf 322}, 256\\
K\"onigl A., 1981, Astrophys. J. {\bf 243}, 700\\
Krymsky G. F., 1977, Dokl. Akad. Nauk SSSR {\bf 234}, 1306, (Engl.
  transl. Sov.
  \hspace*{0.4truecm} Phys.-Dokl. {\bf 23}, 327)\\
Kulkarni S. R. et al., 1998, Nature {\bf 393}, 35\\
Lagage P. O., Cesarsky C., 1983, Astron. Astrophys. {\bf 125}, 249\\
Lieu R., Quenby J. J., Drolias B., Naidu K., 1994, Astrophys. J. {\bf 421}, 211\\
Lucek S. G., Bell A. R., 1994, Mon. Not. R. Astron. Soc. {\bf 268}, 581\\
Medvedev M. V., Loeb A., 1999, Astrophys. J. submitted 
(astro-ph/9904363)\\
Meegan C. A., Fishman G. J., Wilson R. B., Paciesas W. S.,
  Pendleton G. N., Ho-
  \hspace*{0.4truecm} rack J. M., Brock M. N., Kouveliotou C., 1992,
  Nature {\bf 355}, 143\\
Mirabel I. F., Rodriguez, L. F., 1994, Nature, {\bf 371}, 46\\
Naito T., Takahara F., 1995 Mon. Not. R. Astron. Soc. {\bf 275}, 1077\\
Newman P. L., Moussas X., Quenby J. J., Valdes-Galicia J. F.,
  Theodossiou-Ekateri-
  \hspace*{0.4truecm} nidi Z., 1992, Astron. Astrophys.  {\bf 255}, 443\\
Paczy\'nski B., Xu G., 1994, Astrophys. J. {\bf 427}, 708\\
Parker E. N., 1958, Phys. Rev. {\bf 109}, 1328\\
Quenby J. J., Drolias B., 1995, in Proc. 24th Int. Cosmic Ray Conf. {\bf 3},
  p. 261,
  \hspace*{0.4truecm} Rome\\
Quenby J. J., Lieu R., 1989, Nature {\bf 342}, 654\\
Ostrowski M., 1988, Mon. Not. R. Astron. Soc. {\bf 233}, 257\\
Ostrowski M., 1991, Mon. Not. R. Astron. Soc. {\bf 249}, 551\\
Ostrowski M., 1993, Mon. Not. R. Astron. Soc. {\bf 264}, 248\\
Ostrowski M., 1994, Comments Astrophys. {\bf 17}, 207\\
Ostrowski M., Schlickeiser R., 1996, Sol. Phys. {\bf 167}, 381\\
Schatzman E., 1963, Ann. Astrophys. {\bf 26}, 234\\
Takahara F., Terasawa T., 1990, in Proc. ICRR Int. Symp. on {\it
  "Astrophysical
  \hspace*{0.4truecm} Aspects of the Most Energetic Cosmic Rays"},
  Kofu\\
Tingay S. J. et al., 1995, Nature {\bf 374}, 141\\
Vermeulen R. C., Cohen M. H., 1994, Astrophys. J. {\bf 430}, 467\\
Vietri M., 1995, Astrophys. J. {\bf 453}, 883\\
von Montigny C. et al., 1995, Astrophys. J. {\bf 440}, 525\\
Waxman E., 1995, Phys. Rev. Lett. {\bf 75}, 386\\
Waxman E., 1997, Astrophys. J., {\bf 485}, L5\\
Woods E., Loeb A., 1995, Astrophys. J. {\bf 453}, 583

\end{document}